\documentclass[num-refs]{nbdt-article}

\usepackage{siunitx}
\usepackage{ccfonts}
\usepackage{tcolorbox}
\usepackage{xcolor}

\papertype{Original Article}
\paperfield{Neural data science/analysis}

\title{\Large A standardised open science framework for sharing and re-analysing neural data acquired to continuous stimuli}

\author[1]{\normalsize Giovanni M. Di Liberto}
\author[2]{\normalsize Aaron Nidiffer}
\author[3,4]{\normalsize Michael J. Crosse}
\author[5]{\normalsize Nathaniel J. Zuk}
\author[6,7]{\normalsize Stephanie Haro}
\author[8,1]{\normalsize Giorgia Cantisani}
\author[1]{\normalsize Martin M. Winchester}
\author[1]{\normalsize Aoife Igoe}
\author[1]{\normalsize Ross McCrann}
\author[1]{\normalsize Satwik Chandra}
\author[2,9]{\normalsize Edmund C. Lalor}
\author[10]{\normalsize Giacomo Baruzzo}

\affil[1]{School of Computer Science and Statistics, University of Dublin, Trinity College, Ireland; ADAPT Centre, Trinity College Institute of Neuroscience}
\affil[2]{Dept Biomedical Engineering, Dept Neuroscience, Del Monte Institute for Neuroscience, Center for Visual Science, University of Rochester, NY, USA}
\affil[3]{Segotia, Galway, Ireland}
\affil[4]{Department of Mechanical, Manufacturing and Biomedical Engineering, TCBE, Trinity College Dublin, Ireland}
\affil[5]{Department of Psychology, Nottingham Trent University, Nottingham, UK}
\affil[6]{Human Health and Performance Systems, MIT Lincoln Laboratory, Lexington, Massachusetts, USA}
\affil[7]{Speech and Hearing Bioscience and Technology, Harvard Medical School, Boston, Massachusetts, USA}
\affil[8]{Laboratoire des systémes perceptifs, Département d’études cognitives, ENS, PSL University, CNRS, 75005 Paris, France}
\affil[9]{Dept Biomedical Engineering, Center for Visual Science, University of Rochester, NY, USA}
\affil[10]{Department of Information Engineering, University of Padova, Padova, Italy}

\corraddress{College Green, Dublin 2, Ireland}
\corremail{gdiliber@tcd.ie}

\fundinginfo{This research was supported by the Science Foundation Ireland under Grant Agreement No. 13/RC/2106\_P2 at the ADAPT SFI Research Centre at Trinity College Dublin. ADAPT, the SFI Research Centre for AI-Driven Digital Content Technology, is funded by Science Foundation Ireland through the SFI Research Centres Programme. SH was supported by a National Institute of Health (NIH) T32 Trainee Grant No. 5T32DC000038-27 and the National Science Foundation (NSF) Graduate Research Fellowship Program under Grant No. DGE1745303. GC was supported by an Advanced European Research Council grant (NEUME, 787836) }

\runningauthor{Di Liberto et al.}

\begin{document}

\maketitle

\begin{abstract}
Neurophysiology research has demonstrated that it is possible and valuable to investigate sensory processing in scenarios involving continuous sensory streams, such as speech and music. Over the past 10 years or so, novel analytic frameworks combined with the growing participation in data sharing has led to a surge of publicly available datasets involving continuous sensory experiments. However, open science efforts in this domain of research remain scattered, lacking a cohesive set of guidelines. This paper presents an end-to-end open science framework for the storage, analysis, sharing, and re-analysis of neural data recorded during continuous sensory experiments. We propose a data structure that builds on existing custom structures (Continuous-event Neural Data or CND), providing precise naming conventions and data types, as well as a workflow for storing and loading data in the general-purpose BIDS structure. The framework has been designed to interface with existing EEG/MEG analysis toolboxes, such as Eelbrain, NAPLib, MNE, and mTRF-Toolbox. We present guidelines by taking both the user view (rapidly re-analyse existing data) and the experimenter view (store, analyse, and share), making the process straightforward and accessible. Additionally, we introduce a web-based data browser that enables the effortless replication of published results and data re-analysis.
\keywords{EEG, ECoG, MEG, Continuous stimuli, Data sharing, CND, CNSP}

\end{abstract}

\section{Introduction}
In our daily life, we navigate complex sensory environments containing overlapping streams of information, such as auditory, visual, and somatosensory signals. The human brain processes these signals, while integrating and interpreting them based on our prior knowledge and expectations \cite{1}. Decades of impactful discoveries have shed light on the neural architecture of sensory processing through ingenious, well-controlled, laboratory-based experiments \cite{2,3,4}. The use of such controlled methods has led to increased interest in testing whether these findings generalise to tasks involving naturally-occurring sensory signals that occur over long (> 1 second) periods of time – what we refer to as continuous sensory experiments \cite{5,6} – such as listening to continuous speech (e.g., audiobooks, lectures, podcasts), listening to music, or watching videos (e.g., movies, short films). These continuous sensory experiments are designed and analysed differently to those that rely on the presentation of more discrete stimuli, expanding the kinds of questions that can be addressed. For example, would the findings on the neural processing of isolated syllables and words apply to natural speech? \cite{7} Furthermore, tasks involving continuous stimuli enable the study of neural processes that would be otherwise inaccessible, such as the connection between sound statistics and music enjoyment \cite{8}. 

There is a growing body of studies involving continuous speech and music \cite{9,10}, with data and analysis code being shared more frequently within the community \cite{11,12,13,14}. However, in contrast to scientific fields with a more evolved open-science framework (for example, bioinformatics – Box 1), clear open-science guidelines and tools are missing when working with continuous sensory experiments, fragmenting the current literature into lab-specific procedures for data storage, analysis, and sharing. We aim to improve the degree to which results can complement each other by suggesting a framework for continuous human auditory neurophysiology experiments. This paper presents a cohesive end-to-end open science framework, with user-friendly guidelines and tools for the storage, analysis, and sharing of neural data acquired to continuous stimuli, as well as analysis code. We also demonstrate how this framework enables immediate access to existing datasets, offering libraries and tools for the rapid replication of published results, re-analysis (e.g., power analysis) with different configurations, and hypothesis formulation through a new data simulation toolkit. Finally, we present a web-based data browser that enables effortless replication of results and data re-analysis. 

\paragraph{Investigating sensory processing with continuous stimuli}

In neurophysiology, the nature of the sensory stimuli influences the subjective experience of the listener and what sensory processes are recruited in the experiment. Sensory experiments can be categorised in two ways based on the type of stimuli used. Sensory inputs can be presented continuously over a long (several seconds or minutes) period of time, or discretely as a sequence of many short (millisecond) events (Figure 1). Continuous sensory experiments refer to tasks involving uninterrupted auditory, visual, or tactile sensory streams \cite{15,16,12,17,18,19,20,21}. By contrast, discrete sensory experiments involve presentation streams that are based on brief, separate events, such as oddball paradigms involving repetitions of isolated standard/deviant syllables. Music and speech streams are considered continuous, despite being composed of a sequence of well-defined segmental units (i.e., music notes, words). For example, in monophonic melodies, the listener does not perceive the separate disconnected notes, but instead perceives a continuous melody where each note is meaningfully placed within a broader context. Similarly, silence in continuous speech has a communicative role \cite{4,22,23,24,25}, as opposed to silence that is artificially introduced to separate the repetition of syllables or words in discrete sensory experiments. It should be noted that one can extract discrete features from continuously presented sensory streams. Some experiments are continuous in their presentation of the stimuli, but are designed to investigate the neural processing of both continuous and discrete features embedded in continuous streams (e.g., the sentence onsets of continuous speech; switching of attention in multi-talker scenarios; Figure 1). In this work, continuous sensory experiment refers to how the temporal nature of the task is perceived by the participants. The analysis that follows may involve the examination of continuous, discrete, or mixed features, providing us with experimental data to study cognition in the context of tasks that are close to what is routinely experienced in the real-world (e.g., listening to music through headphones, listening to the radio).

The temporal nature of the experiment (i.e., discrete vs. continuous) can change how the brain processes the input7 and determine what analyses techniques are most appropriate to characterise these processes. In experiments involving discrete events, the only piece of stimulus-related information necessary are the timestamps indicating the onset of each event (e.g., syllable onsets, button-click) and the categorical labels indicating how to classify each event (e.g., frequent vs. infrequent, syllable identity). This information can be used to conduct the well-known Event-Related Potential/Field analysis (ERP/ERF). Timestamps are used to segment neural data (e.g., EEG/MEG) or other physiological data (e.g., pupillometry) into epochs, and the categorical labels are used to group those epochs according to their experimental condition. Conversely, the absence of carefully designed discrete events makes continuous sensory experiments less suitable for ERP analyses (although see Khalighinejad et al. \cite{26}), while system identification analyses involving the estimation of input-output mapping are more appropriate. This type of analysis has typically been conducted using continuously varying properties of the stimulus, such as the speech envelope, but it can also be carried out using event timestamps, which are represented using a binary mask i.e., sequences of zeros and ones, where the latter indicate the onset or the entire occurrence of a given event (e.g., word onset). Note that some properties are represented in a manner that is discrete in time but continuous in magnitude, such as semantic dissimilarity and lexical surprise \cite{27,28,29}.

\begin{figure}[bt]
\centering
\includegraphics[height=3cm]{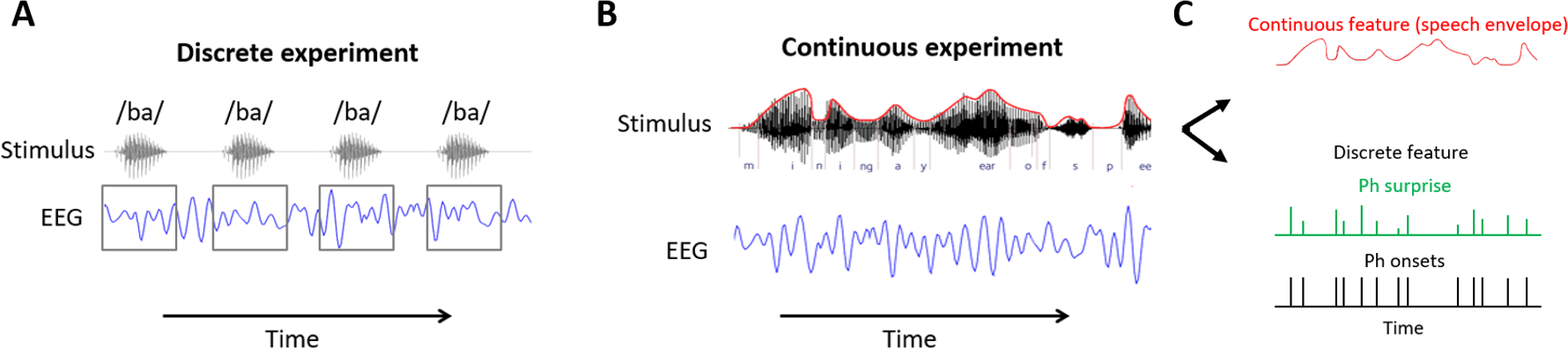}
\caption{\textbf{Investigating sensory processing with discrete versus continuous stimuli.} (A) An EEG experiment involving listening to individual syllables that are presented sequentially. The discrete syllable timestamps can be used to epoch the data. (B) An EEG experiment involving natural speech listening. The stimulus was continuous speech, hence there is no isolated, discrete event in the traditional experimental sense that can be used to epoch the data. (C) A continuous experimental stream is rich in information and can be described using both continuous and discrete features. Some stimulus features are encoded continuously by the brain so the feature of interest is continuous in both time and magnitude, such as the sound envelope (top panel). Events of interest in the continuous stream can be described using discrete timestamps. For example, phonemic surprisal describes the level of unexpectedness of each phoneme based on the preceding context (middle panel). Additionally, a binary mask can be used to indicate discrete events over the continuous experiment, such as phoneme onsets (third panel).}
\end{figure}

When considering tasks such as speech and music listening, continuous stimuli enable the design of more naturalistic experiments. But why is it so important to study sensory processing with naturalistic tasks? One reason is to test whether findings from artificial paradigms generalise to brain processes during more natural, real-world conditions. Secondly, more naturalistic scenarios can increase participant comfort and engagement, allowing testing of individuals who typically struggle with traditional experiments (e.g., children, individuals with neurocognitive deficits). Thirdly, continuous sensory tasks – such as speech listening – engage many neural processes, allowing us to assess them and their interaction simultaneously, rather than studying those processes in isolation (e.g., syllables in the context of a sentence vs. a syllable oddball paradigm). Finally, there is one additional advantage that does not necessarily impact the experimenter, but rather the research community at large. Namely, while neurophysiology experiments may be designed to tackle one specific hypotheses, data sharing may enable both results replication and the re-analysis of the data to study different research questions. More naturalistic experiments involving continuous stimuli can be reanalysed in numerous ways with different objectives (e.g., Natural Speech dataset \cite{11}), as opposed to more traditional paradigms involving short or discrete stimuli, where the number of possible re-analyses is limited. Re-analyses of this kind can be particularly useful for methodological development, as well as for hypothesis design and preliminary results generation, which would then be followed by further testing on other new or existing datasets. Indeed, a word of caution is necessary here, as this new possibility comes with risks such as “p-hacking” or, similarly, “fishing” for results. Nevertheless, the tools for mitigating such risks exist and should be adopted, such as the careful use of cross-validation within and between datasets, and by replicating the results on a new dataset. 

\definecolor{mygreen}{RGB}{200, 255, 200}
\tcbset{colback=mygreen, coltext=black, colframe=green!70!black, boxrule=0.5mm, arc=4mm}

\begin{tcolorbox}
\textbf{Box 1: Data sharing in Bioinformatics: What have we learnt?}

Is data sharing worth it? Yes, when it’s done right! The field of Bioinformatics has implemented an open science framework since the very beginning. In a way, the field itself is a large open science initiative. The scientific publication pipeline includes strict rules for data and code sharing, which is similar but much more rigorous and standardised than in neuroscience at present.

In terms of data, the deposition of omics data in public community repositories is mandated by most funding agencies and journals. Data deposition includes the release in public repositories of raw data, meta data, and (optionally) preprocessed data. Public repositories offer guided procedures that help authors upload both data and metadata, requiring the adoption of specific file formats, and providing both automatic and manual checks of the validity of the uploaded data. Data released in such public repositories obtain a persistent identifier, and can be queried and downloaded by other users using dedicated APIs or through an interactive web interface. The effort of managing such large public repositories is typically handled by the collaborative support of several national, international and interoperative research agencies, such as National Center for Biotechnology Information (NCBI), National Institutes of Health (NIH) and European Bioinformatics Institute (EMBL-EBI).

In terms of code, the release of source code in public code repositories is mandatory for most journals, as well as the specification of the software version and the release of the chunks of code used to obtain the main results/figures. For example, one of the leading methodological journals in the field, i.e. Bioinformatics, requires that authors provide a self-contained and easy-to-use implementation of the developed software together with test data and instructions on how to install and run the software. Software source code must be freely available on a stable URL, such as GitHub. Both submitted software version and test data must also be archived on dedicate repositories, such as Zenodo.

In summary, while this may mean more is required at the time of publication submission, this additional effort has an invaluable positive impact on scientific research, especially on transparency, result replication, and data re-analysis. 
\end{tcolorbox}

\begin{figure}[bt]
\centering
\includegraphics[height=4cm]{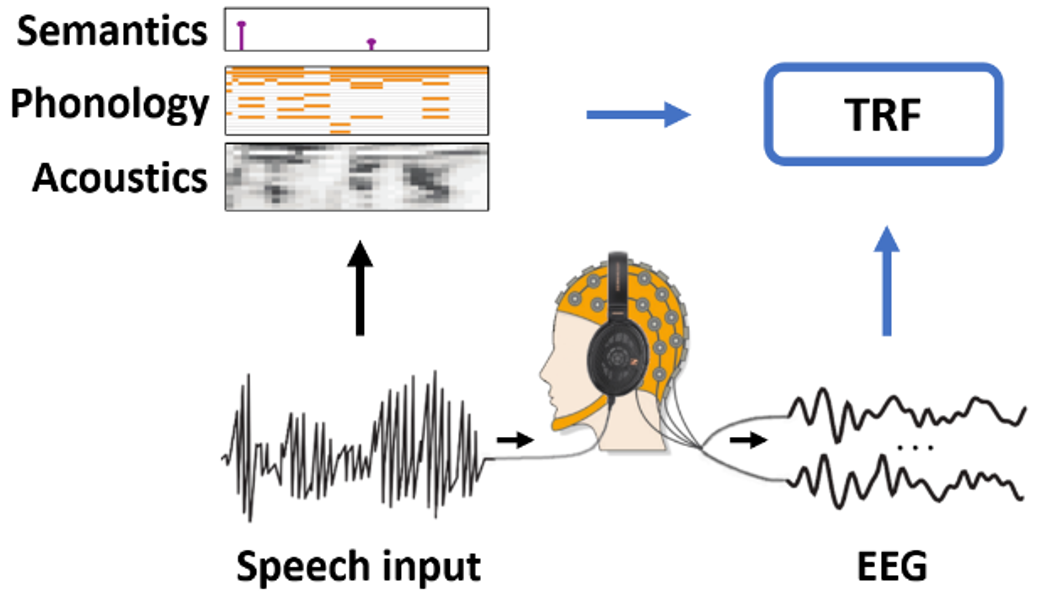}
\caption{\textbf{The multivariate temporal response function (TRF) framework}. Neural signals are recorded as participants are presented with sensory stimuli, such as speech. A specific task may be part of the experiment (e.g., answering comprehension questions). Speech and language features are extracted from the stimulus and are simultaneously related to the neural signal with methods such as ridge regression. The weights of the model inform us about the spatiotemporal relationship between each stimulus feature and the neural signal.}
\end{figure}

\paragraph{Processing neural data acquired to continuous sensory stimuli.}

The past few years have seen a surge in the use of naturalistic tasks, largely due to the recent development of neural signal analysis methods for continuous sensory experiments. This report involves analytical frameworks such as forward models \cite{20,30}, backward models \cite{30}, and canonical correlation analysis (CCA; Figure 3C) \cite{31}, with a focus on forward and backward models, the former also referred to as Temporal Response Functions (TRF), which capture how neural signals respond to changes in a specific sensory feature (Figure 2). The TRF is a quantitative estimate of the stimulus-response relationship, assuming approximate linearity and time-invariance. TRF analyses rapidly gained popularity primarily due to the fact they can be used to study how our brains track, encode, and build expectations of continuous stimuli, such as speech and music. For example, the envelope TRF estimates the relationship between the sound envelope, a key property of speech processing \cite{32,33}, and the neural response as measured using EEG/MEG recordings for example. For implementation, we refer to the mTRF-Toolbox \cite{23}, a library for estimating multivariate TRFs based on regularised linear (ridge) regression. Please see Crosse et al., \cite{30,35} for detailed information on the multivariate TRF methodology and interpretation and Obleser and Kayser \cite{36} for a perspective on the neural mechanisms that may generate speech TRF results. Of course, alternative approaches exist for relating neural data to ongoing stimuli, such as CCA \cite{31}, or even other types of neural recordings, such as human and non-human intracranial EEG, EOG, ECG, galvanic skin response, and pupillometry, which will not be discussed here for simplicity. More recently, the substantial leap in machine learning methodologies, especially deep learning models (e.g., GPT-2, Mistral, Music transformer), have provided further opportunities for augmenting the set of stimulus descriptors when studying the neural processing of continuous stimuli. In particular, such large language models can be related to neural data directly by considering the weights in their hidden layers \cite{37}, or they can be used to estimate specific linguistic aspects of a speech stream, such as lexical surprisal, which can then be related to neural responses with standard TRF methods \cite{29}.

\begin{figure}[bt]
\centering
\includegraphics[height=8cm]{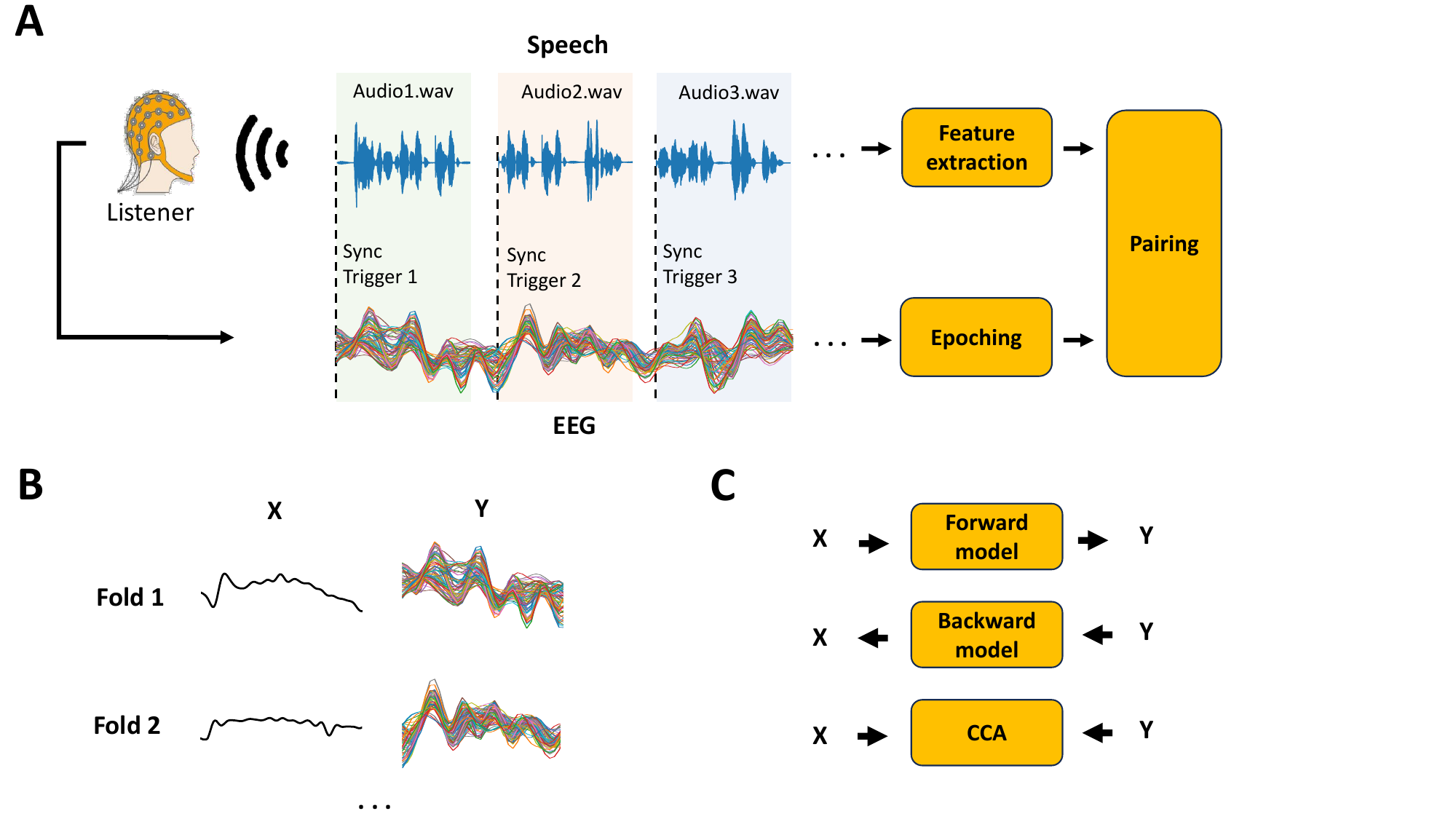}
\caption{\textbf{Continuous sensory experiment data acquisition and analysis.} The figure focuses on a typical continuous speech listening scenario, where X represents the stimulus feature of interest and Y the preprocessed neural signal. (A) EEG/MEG is recorded as the participant listens to speech segments. Synchronisation triggers are used to epoch data into trials of continuous speech, in contrast to being used to epoch data around discrete stimulus tokens. The value of each trigger corresponds to the index of the audio file (e.g., 1: audio1.wav, 2: audio2.wav). Stimulus features (e.g., sound envelope) are extracted for every audio-file and paired with each corresponding EEG/MEG epoch. (B) Neural data and stimulus feature pairs are organised into data structures, X and Y, that are time synchronized. Trials of continuous responses can be used as the folds for a leave-one-fold-out cross-validation. (C) X and Y can be used to investigate the EEG/MEG encoding of the stimulus features of interest using a forward model (TRF) \cite{38} approach. Conversely, X and Y can be used to build decoding or backward models \cite{39}. While mTRF-based forward and backward models are limited to multivariate-to-univariate mappings, relationships where both X and Y are multivariate can be studied with methods like canonical-correlation analysis (CCA) \cite{31}.}
\end{figure}

Given the importance and power of experiments based on continuous, naturalistic stimuli, there has been a marked increase in the number of available datasets recorded using such stimuli. However, these datasets have been somewhat underexploited given the lack of any coherent set of data storage and sharing protocols across the community. This has meant that data are stored in ways that are particular to each research group, or even inconsistent within the same group – making it inefficient and sometimes impossible for other researchers to use those data to answer new or complementary questions. Nonetheless, while these inconsistencies extend to aspects such as data formatting, naming of variables and files, and the fine details of the analysis, there exist strong procedural similarities that can be codified. Figure 3 attempts to depict these consistent procedural steps by considering a widely used and particularly simple scenario involving listening to natural speech segments (e.g., chapters of an audio-book). In that case, the first part consists of a) extracting the features of interest from the stimulus (e.g., speech envelope); b) segmenting the neural data (epoching); and c) resampling and aligning the speech features with the corresponding neural segments (pairing). The resulting stimulus features and neural segments are precisely synchronised (same start sample and number of samples). The combination of all stimulus and neural signal segments, which we refer to as X and Y respectively, can then be used to fit input-output models, such as TRFs and CCA. This manuscript also discusses how the key steps in obtaining X and Y, which are depicted in Figure 3, can also be adapted to scenarios involving multiple experimental tasks or conditions (e.g., listening vs. imagery).

\section{Results}

\paragraph{The proposed end-to-end framework}

Here, we present end-to-end specifications and resources for the analysis of continuous sensory event neural data, including data storage specifications, standardised datasets, learning resources, and analysis tools. One of the key motivations of having such a cohesive data management procedure is to simplify and expedite all operations from data storage to analysis. To this end, we define a domain-specific data structure, called Continuous-event Neural Data (CND), providing a precise naming convention, folder structure, and data types for data storage in a manner similar to some existing datasets in the literature \cite{11,40,41,42,43}, offering well-defined how-to guidelines on storing new data. This will enable the immediate use of all resources produced by this project, from preprocessing scripts, analysis scripts and GUIs for running TRFs and CCA analyses, as well as a simulation toolkit. CND can be loaded directly by other widely used toolkits such as NAPLib \cite{44} and Eelbrain \cite{45}. Furthermore, we provide import/export functions connecting CND with the widely-used Brain Imaging Data Structure (BIDS), allowing users to store and share those data in a standardised manner, and consequently to have access to general-purpose toolkits for neural signal analysis (e.g., MNE \cite{46}, EEGLAB \cite{47}). This work has been carried out by an interdisciplinary, international team that was initially assembled in 2020 with the goal of propelling open science and thus supporting basic and translational research in this domain. The resulting open science initiative, named Cognition and Natural Sensory Processing (CNSP), is an ongoing international research collaboration with headquarters in Trinity College Dublin. 

\begin{figure}[bt]
\centering
\includegraphics[height=6cm, trim={0 {5.5cm} 0 0}, clip]{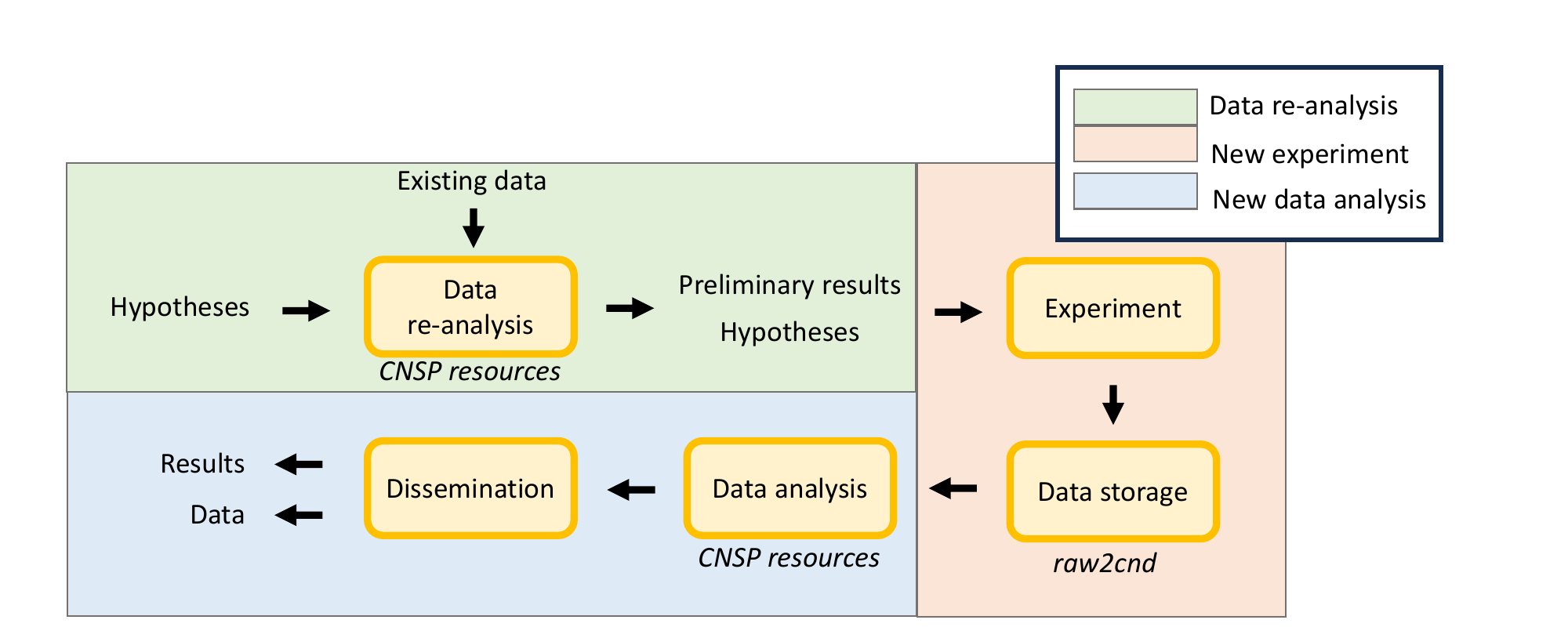}
\caption{\textbf{An end-to-end open science framework for hypothesis formulation and testing.} Hypotheses are preliminarily tested and refined by re-analysing existing data. This step can either be carried out by using a publicly available dataset that is similar to the target scenario, or via data simulation. The refined hypotheses are then tested on a new experiment. Data stored in a standardised format (e.g., CND) can then be analysed with the same procedure used with the preliminary data. The proposed standardised data structure connects these steps into a cohesive end-to-end framework, enabling the immediate use of the resources in this project, such as analysis tools and simulation toolkit.}
\end{figure}

This report is organised from the viewpoint of an experimenter that is conducting a new neurophysiology study from scratch and intends to formulate precise hypotheses supported by existing data, and then run a new experiment (Figure 4). First, we present the CND standard for data storage (Storage and sharing). Second, we detail a standardised analysis pipeline for neural data analysis, from minimal preprocessing, TRF, CCA, and cross-correlation analyses, including libraries and learning resources such as video-tutorials (Analysis pipeline). Third, a new data browser is presented that allows for rapid result replication and re-analysis of previous datasets, as well as the performance of comparisons between datasets, models, and methodologies; all with a user-friendly environment (Data browser GUI). Next, we present a simulation toolkit and describe its value for experimental design and hypothesis formulation (Simulating). Finally, we discuss the limitations of the current framework, ongoing work, and future directions, including pointing at areas of development where the research community is encouraged to contribute (Limitations and future directions).

\begin{tcolorbox}
\textbf{Box 2: Open-science resources used in this report}

\begin{itemize}
    \item The online documentation consists of datasets, analysis pipelines, libraries, tutorial scripts, video-tutorials, video-lectures, and example scripts which are publicly available here https://cnsp-resources.readthedocs.io and on the Cognition and Natural Sensory Processing (CNSP) Workshop website https://cnspworkshop.net
    \item The code for the CNSP data browser, including the scripts used for neural data preprocessing, TRF model estimation, stimulus feature extraction, and simulation toolkit are available on the CNSP website and on the CNSP GitHub page (https://github.com/CNSP-Workshop)
    \item The mTRF-Toolbox is used for demonstrating TRF analyses (https://github.com/mickcrosse/mTRF-Toolbox), but the same considerations apply to other compatible toolboxes, such as Eelbrain (https://github.com/christianbrodbeck/Eelbrain) and NAPLib (https://github.com/naplab/naplib-python)
    \item All datasets were stored according to the Continuous-event Neural Data (CND) format, whose specifications are also available on the CNSP Initiative website (https://cnspworkshop.net/cndFormat.html). To date, CND datasets include tasks such as speech listening, auditory attention, music listening and imagery, and a longitudinal dataset on nursery rhyme listening in infants
\end{itemize}
 
\end{tcolorbox}

\paragraph{Storage and sharing}

Sharing data and analysis scripts is key for advancing science and promoting transparency in research. However, the lack of clear guidelines can lead to a frustratingly heterogeneous set of publicly-available resources. Shared datasets and scripts in the field of auditory neuroscience are often difficult to use without additional help from the authors. In some cases, the authors may no longer be available or may not remember how to run their own code, which can further hinder the process. On the other hand, overly rigid guidelines for data sharing can also be problematic, as they can make the sharing process overly complicated and become counterproductive. The framework described here aims to strike a balance between these two extremes, providing guidelines that are specific enough to be useful but not so complex as to be burdensome. Unlike general-purpose tools such as EEGLAB and MNE, our objectives are domain-specific, as they target continuous sensory experiments only and were developed by focusing on the growing audio and audio-visual literature on speech and music perception. This allowed us to design guidelines that keep the methods as straightforward as possible for both experimenters (who collect and store their data) and future users (who load and re-analyse such data). 

Firstly, experimenters should ensure that they have the right to share all data, including the original stimulus files (e.g., audio files); and this should be verified before data collection, at the experimental design stage. This trivial yet sometimes neglected point is key to avoiding a reliance on the “available upon request” statement, which translates to multiple time-consuming direct interactions between users and those who shared the data. Secondly, it is extremely important that experimenters adopt a standardised data structure. Here, standardised refers to the presence of a clear specifications on data types, folder structure, and naming convention, with documentation and how-to guidelines. Crucially, it also indicates the existence of import/export scripts allowing for the effortless conversion of the stored data from one standardised data structure to another. As such, the experimenter can adopt the data structure that best fits their project and expertise, and then export to a different data structure when needed.

The data structure that is most widely adopted for data sharing in neuroimaging and, more recently, in neurophysiology is BIDS. This data structure is general-purpose, as it can be used to store virtually any kind of experimental assets or data involving fMRI, EEG, MEG, and other recording modalities. BIDS and its extensions are also compatible with widely used toolboxes such as EEGLAB (through a specific plugin) and MNE (via the extension MNE-BIDS), supporting rigorous science by providing tested analysis scripts and pipelines, reducing the risk for analysis errors due to bugs in the code, for example. To fulfil the goal of being general-purpose, BIDS focuses on storing raw data, while it leaves unspecified constraints that may be necessary in particular domains of research. The input-output models used in continuous sensory neural data analysis require the extraction of stimulus features and the alignment of such features and the corresponding neural responses, as depicted in Figure 3. However, the lack of domain-specific detailed specifications on how experimenters should store their stimuli (e.g., raw audio files or midi files, sound envelopes, note onsets), has led to disparate interpretations that have to be addressed in each specific toolbox or custom script. 

While BIDS can indeed be used for saving continuous sensory neural data, the literature lacks clarity on how that should be done. Based on existing custom formats for EEG data storage, we define a domain-specific data structure that is tailored to continuous sensory experiments, called the Continuous-event Neural Data (CND) structure. This data structure provides a more tailored approach for data storage and sharing, aiming to simplifying result replication and re-analysis, as input-output models such as TRF and CCA can be run directly on the CND structure. The domain-specific constraints and guidelines ensure consistency across datasets, as well as simplifying the data storage procedure. Crucially, we connected CND and BIDS via specific import/export functions, enabling the use of general-purpose toolkits, where useful. In doing so, we provide a pipeline connecting existing data storage approaches \cite{11,40,41,42,43} with BIDS, as well as processing resources that can promptly be applied for results replication and data analysis. Next, we present a summary of the CND data structure. Please also refer to the CNSP website (https://cnspinitiative.net) and CNSP documentation (https://cnsp-resources.readthedocs.io) for more detailed specifications.

\subparagraph{Continuous-event Neural Data (CND)} 

CND is a domain-specific data structure tailored to continuous sensory neurophysiology experiments. The data structure was designed to feed directly into input-output analysis methods, such as encoding models (e.g., TRFs), decoding models, and CCA. To so do, CND stores input (e.g., speech features) and output (e.g., neural responses) as numerical matrices, already segmented and synchronised (i.e., the first sample in both the input and output segments correspond to the same timestamp). In contrast to BIDS, CND includes detailed guidelines on how to store a data structure with the stimulus features (e.g., the sound envelope). In this way, the features that are key to a given study are already included in the data structure and are readily available for analysis, while future users will additionally have access to the raw data for their preferred preprocessing pipeline. The CND folder structure also includes subfolders for the raw stimulus and neural data files. Figure 5A depicts the folder structure, with folder and file names that should be used exactly as reported here (see Methods). This will ensure compatibility with CNSP resources such as the tutorials, Data Browser, and Simulation Toolkit. Note that the current version of the CNSP resources (v1.0) utilises MATLAB data structures for storing the CND data, as most of the datasets that are currently part of the resources were part of studies initially analysed in the MATLAB programming language. While the same was also true for seminal analysis toolboxes such as NoiseTools \cite{48} and mTRF-Toolbox \cite{23}, some have now been extended to support Python too, thanks to the release of packages such as mTRFpy \cite{49}. Regarding what format to store neural data in, while the current support is limited to .mat files, there is no set limitation in that regard, as the concept in Figure 5 can (and will) be implemented in other storage formats – with the caveat that a dedicated loading function should be added to the libraries for each new format. At present, the proof-of-concept is the ability of loading CND from Eelbrain and NAPLib, which are both Python-based and, in principle, could be used to store the data in any other useful format.

Please refer to the online documentation for guidelines on setting up a new CND data folder and with further details on what should (and should not) be included in each of the sub-folders in Figure 5 (https://cnsp-resources.
readthedocs.io/en/latest/cndPage.html). Please refer to Figure 5B for the exact naming convention (i.e., stim.data contains the various stimulus features, stim.names contains the name chosen to describe and identify each feature).

\begin{figure}[bt]
\centering
\includegraphics[height=9cm]{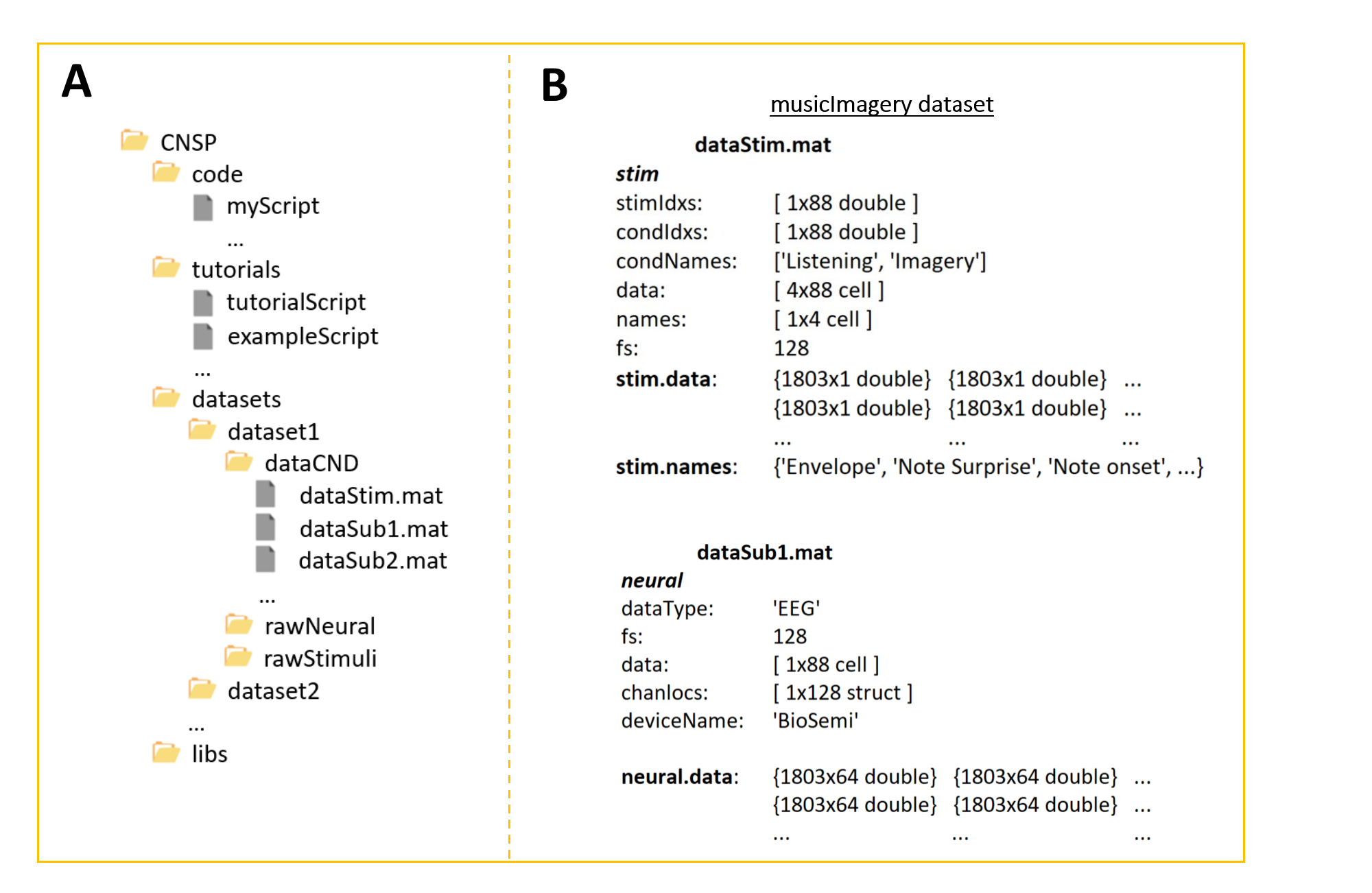}
\caption{\textbf{The Continuous-event Neural Data (CND) structure.} (A) CND folder structure. ‘dataStim.mat’ refers to the stimulus features, while ‘dataSubX.mat’ refers to the neural data. In the example, all participants were presented with the same stimuli. It is also possible to associate a ‘dataStim’ file to each participant if needed e.g., dataStim1.mat, dataStim2.mat. (B) Example of dataStim and dataSub structures. dataStim contains the stimulus data matrices (stim.data) for each of the four stimulus features and for each of the 88 experimental trials. ‘names’ specified what the four stimulus features are i.e., envelope, note expectation, note onset, metronome. ‘stimIdxs’ indicate the stimulus index of each trial. The ‘rawStimuli’ folder should include documentation regarding what stimulus file corresponds to each index. Here, indices 1–4 refer to musical pieces named ‘chor-096’, ‘chor-038’, ‘chor-101’, and ‘chor-019’ respectively. ‘condIdxs’ refers to the task that was carried out in each trial i.e., music listening or imagery. The EEG data structure is straightforward in that it simply contains basic meta-data, such as the ‘deviceName’, and a data matrix (neural.data), with a [samples x channel] data matrix per trial. }
\end{figure}

\subparagraph{Interoperability: connecting CND with existing resources}

In the last decade or so, the importance of adopting open science strategies in our field has progressively become clearer. As a result, several lab-specific toolboxes have been released, each based on custom code written by individual scientists for their own work (e.g., NoiseTools \cite{48}, mTRF-Toolbox \cite{30,35}, Eelbrain-toolkit \cite{45}, NAPLib \cite{44}, Unfold-toolbox \cite{51}). In some cases, distinct toolkits run similar (but not identical) analyses. Therefore, they cannot be easily compared because they are not directly compatible. Interconnecting these tools can be time-consuming for scientists skilled in computer programming and inaccessible to others. As such, one key output of the present work is that we have established an infrastructure to connect such resources. Specifically, our approach consists of replacing lab-specific custom data structures with CND, while BIDS serves as a go between, connecting this work with general-purpose analyses resources such as MNE-Python (Fig. 2C) \cite{46}. Finally, domain-specific toolboxes such as Eelbrain and NAPLib have also been extended by their respective authors to support the CND data structure.

\begin{tcolorbox}
\textbf{Box 3: Storing raw data in CND format: practical considerations}

Here are the key practical steps for building a new project folder structure:
\begin{itemize}
    \item Create a project folder on your hard-drive (e.g., \$ROOT/studentProject1/).
    \item Copy the content of the CNSP GitHub folder into your new project folder (https://github.com/CNSP-Workshop/CNSP-resources/tree/main/CNSP). As a result, you will see the essential subfolders ‘datasets’ and ‘libs’ in your project. While the ‘dataset’ folder is initially empty, it is meant to store all the datasets that are part of a given project.
    \item Use the subfolder 'code' for the analysis code specific to your project. Note that you can also access the ‘exampleScripts’ folder, which contains additional code that might serve as a blueprint for your analysis.
    \item The ‘datasets’ folder contains a ‘datasetTemplate’ subfolder (e.g., \$ROOT/studentProject1/datasets/ datasetTemplate/). Copy and rename this folder with the name of the experiment. The folder contains three subfolders, for the raw neural data (e.g., 'rawNeural'), raw audio stimuli (e.g., 'rawSoundStimuli'), and CND dataset ('CND'). Copy your raw experiment data in the corresponding folders. Repeat for all datasets in your project. Note that the CND dataset can potentially capture various types of data, from EEG, MEG, and intracranial EEG, which were the focus of this manuscript, to EOG, heartbeat, galvanic skin response, and pupillometry for example.
    \item Convert your raw data to CND, saving the dataStim and dataSub into the 'CND' subfolder. We provide guidelines, a video-tutorial, as well as a MATLAB/Octave scripts for converting EEG BioSemi datasets into CND (bdf2cnd.m) and a python script for converting BIDS datasets into CND (bids2cnd.py). Both files are available on the documentation webpage (https://cnsp-resources.readthedocs.io/en/latest/cndPage.html). Note that the pipeline is experiment and device specific, so the script will have to be modified to fit the specific neurophysiology recording device and synchronisation trigger protocol (see Methods). Indeed, we encourage researchers to contribute by sharing a conversion script for other devices and technologies.
    \item You are good to go! You can now run the tutorial code or the GUI to process your CND data.
    \item Storing the raw data might lead to very large archives. We suggest sharing both the original version of the dataset, with the raw data, and a lite version of the dataset, only with the CND data. Regarding the original version of the dataset, we strongly encourage to also share the scripts using for extracting and converting to CND the stimulus features when possible, as future users might be interested in extracting a different set of features. This will enable future users to do so while ensuring correct synchronisation between stimulus features and neural signal.
\end{itemize}
\end{tcolorbox}

\paragraph{Analysis pipeline}

One of the core goals of the CNSP Initiative is to facilitate analysis, sharing, and reanalysis of continuous sensory event neural data. To this end, a library of scripts has been shared with the CNSP community as part of a workshop series. These scripts, which include preprocessing, analysis, and plotting routines as well as numerous low-level support functions (e.g., filtering, down-sampling) were intended to serve as a blueprint for standardising future analysis pipelines. A standardised approach allows for easy sharing, reanalysis, and comparison across datasets and methodologies. We don’t see these scripts as a finished product. Instead, we encourage users to contribute additions or corrections through the CNSP repository on GitHub (https://github.com/CNSP-Workshop/CNSP-resources) \cite{52}. These scripts were written to require minimal or no customisation when analysing a new dataset in CND format. As such, they allow for easy interfacing with many toolboxes (such as the mTRF-Toolbox \cite{30,35}) used for the analysis of continuous neural signals. This is possible by restricting the domain of interest to continuous sensory listening scenarios, or other scenarios that can be coded similarly. For example, although not optimal, a typical mismatch-negativity scenario could be stored and analysed according to these same guidelines.

\subparagraph{Preprocessing}

One benefit of analysing neural data with methodologies such as forward models, backward models, or CCA is that the analysis often only requires minimal preprocessing if the dataset is not excessively noisy. For a typical experiment, preprocessing involves at least a high-pass filter or detrending step to remove potential drift, down-sampling data to a more manageable size, and epoching the neural recording into manageable segments \cite{35}. These preprocessing steps are included in the shared CNSP libraries and example scripts. Specifically, they include (in this order) low-pass filtering (Zero-phase shift Butterworth low-pass and high-pass filters of order 2 are used as a standard choice, but others should also be considered), down-sampling, high-pass filtering, bad-channel detection and replacement via a spline interpolation, and data re-referencing (e.g., global average re-refencing, mastoid re-refencing). These are also the typical preprocessing steps in numerous previous papers in the relevant literature \cite{15,44,53,54,55}, which have been discussed in detail in the context of TRFs by Crosse et al. (2021) \cite{35}. Other work has also successfully utilised that same pipeline for other related methodologies, such as CCA \cite{53}. In addition to these typical preprocessing steps, previous work has also employed a variety of additional methodologies for further cleaning data. For example, EEG and MEG researchers often avail of ICA, DSS, Wiener filters, and other methodologies that can help disentangling the neural activity of interest from unwanted components of the signal \cite{56,57,58,59}. 

Sharing raw data alongside a script for carrying out the preprocessing steps in the related publication is preferable. Nonetheless, the storage memory requirement might be excessive for both user and host in some cases (imagine having to download 1TB of data just to re-run the analysis just to see the impact of changing a given parameter, like filter frequency cut-off or re-referencing channel). So, a good approach that we recommend consists of sharing, in addition to the raw data and preprocessing scripts, a “lite” minimally processed version of the dataset that can be more easily accessible, for example after applying a low-pass filter and down-sampling to reduce the storage memory requirements. Sharing raw data enables secondary users to carry out their own preprocessing, minimising the risk that potentially informative aspects of the data are removed or altered by extra preprocessing steps (e.g., dimensionality reduction, filtering) and reducing potential sources of contamination (e.g., filtering \cite{60}. Sharing minimally processed data also gives secondary users an option to rapidly access the dataset without excessive memory requirements, while also giving them the possibility of carrying out further preprocessing, as opposed to sharing fully preprocessed data, which would make different datasets not comparable due to the use of different preprocessing pipelines. In fact, the minimalist, natural structure of many continuous-events stimulus paradigms (e.g., participant listens to an excerpt of narrative speech) makes replication of findings across different datasets highly convenient so long as there is some consistency in preprocessing. Indeed, when sharing raw or minimally preprocessed data, users have maximum freedom to make choices and include additional preprocessing steps to suit the needs of their analysis pipeline.

\subparagraph{Analysis}

The CND specifications were designed so that data can be easily input into several toolboxes (mTRF-Toolbox, Eelbrain-toolbox, NAPLib) with very little or no reshaping. The mTRF-Toolbox follows the same format and thus no changes are necessary for it, and several conversion functions have already been written to facilitate loading data saved in CND format into the users’ preferred toolbox format. Once data have been loaded, we advocate for a standardised general approach to fitting and testing temporal response functions. Broadly speaking, TRFs should be cross-validated \cite{35}. In other words, evaluation should be done on data unseen by the fitting procedure. This is most typically accomplished by splitting data into training and testing partitions, fitting TRFs on the former partition, and evaluating their ability to predict data from the withheld partition. More conservatively, when optimising the regularisation parameter, we recommend carrying out a nested loop cross validation – that is, three partitions: one for fitting the model, the second one for determining the optimal value for the regularisation parameter, and a third partition for evaluating the final model \cite{35,53}. As for the preprocessing, the TRF analysis pipeline was described in detail by Crosse et al. (2021) \cite{35} and has been applied in several previous studies in the context of, for example, speech listening \cite{15,55} and music listening \cite{12,61}, in both humans and primates \cite{43}. TRF results can be visualised with the functions in the tutorials and example scripts, showing regularisation tuning curves, temporal response functions, and scalp maps. Again, we look to the community to contribute new plotting functionality via the git repository.

\subparagraph{Tutorials and example scripts}

The CND data format was developed alongside the CNSP workshop. Several video-tutorials and example scripts have also been developed that read in data in CND format and implement the abovementioned analysis pipeline. These scripts and video recordings of the tutorial sessions have been shared on the CNSP Github page (https://github.com/CNSP-Workshop/CNSP-resources/tree/main/CNSP/exampleScripts), website (https: //cnspworkshop.net/resources.html), and on the documentation webpage (https://cnsp-resources.readthedocs.io/) to serve as a blueprint for novice users wishing to utilise these resources and use the CND format. They additionally serve to standardise analyses and reduce the risk of coding errors. We don’t see these scripts as a finished product, but as a solid starting point. We encourage members of the CNSP community to contribute changes and corrections so that we can reach a broader consensus on a standard approach to analysing neural data recorded in response to continuous stimuli presentation.

\paragraph{Data browser GUI}

The tutorials and example scripts described in the previous sections serve both as a learning resource and as a blueprint for conducting new analyses. Despite our recommendations on the analysis pipeline, there remain a number of analytic choices that depend on the specific assumptions and objectives of a study. For example, preprocessing choices involving data filtering, as well as analytic choices on the specific methodology (e.g., mTRF, cross-correlation) and parameters (e.g., TRF lag-window) to use. To simplify that process and guide the user through these choices, we develop a graphical user interface (GUI) with MATLAB-software where those specific options are clearly presented to the user. This GUI (https://cnspworkshop.net/resources.html), which is referred to as the CNSP Data Browser, enables the rapid re-analysis of existing datasets, guiding the users through parameter selection. We expect this GUI to be particularly useful when getting familiar with the CNSP resources for the first time or when teaching, as well as serving as a tool for expert users when needing to rapidly re-analyse existing data. While all MATLAB code is shared, we have also made available an executable version of the Data Browser, bypassing the time-consuming process of setting up the libraries and dependencies and which, in fact, does not require an installation of MATLAB software in the first place. 

The Data Browser is a front end for the functions that are already present in the CNSP resources and are used in the tutorials, which are all shared in the project GitHub page (https://github.com/CNSP-Workshop). In addition to the import/export, preprocessing, and analysis functions presented in the previous sections, the Data Browser utilises functions for running power analyses, extracting and combining stimulus features, and simulating neural data. Power analyses are useful for determining an appropriate sample size when planning an experiment. While that analysis can be carried out on previous data from the experimenter or based on published results, new studies may involve unexplored analyses or parameters that could be run on existing data but would typically be heavily time-consuming. The CNSP Data Browser provides access to existing datasets, with rapid re-analysis functions and a simulation toolkit that can be used to generate results which are as close as possible to the new experiment that the user intends to run. In doing so, the Data Browser firstly aims to facilitate the design of future experiments, for example by speeding up the estimation of a reasonable sample size for the future experiment through the functions for power analysis in the GUI – the MATLAB-Software function sampsizepwr was used for implementing the power analysis. Second, it provides a preliminary set of functions for the extraction of basic acoustic features i.e., broadband envelope and spectrogram. Multiple methods of envelope and spectrogram extraction are proposed, based on the Hilbert function and the envelope extraction function from the mTRF-Toolbox. A function is also provided for generating multivariate features by combining existing features (e.g., spectrogram concatenated with lexical surprisal), simplifying the use of multivariate TRF analyses. Finally, the Data Browser is also the front end of the new functions for neural data simulation, which is discussed in detail in the next section. 

Figure 6A depicts the appearance of the Data Browser GUI and Figure 6B shows the result of a forward TRF analysis on the Natural Speech Dataset (audiobook listening task)11 and the diliBach Dataset (Bach monophonic melody listening task) \cite{40}, which replicates the TRF results that can be found in the corresponding papers \cite{12,15}. Here we discuss the functions that are most commonly used: the preprocessing and TRF analyses. For further details please refer to the CNSP website (https://cnspworkshop.net), where documentation, tutorials, and video-tutorials are available and up-to-date. The GUI offers basic preprocessing functions: Data re-referencing (e.g., global average referencing), band-pass filtering, envelope extraction (e.g., for high-gamma or spiking data), downsampling, and bad-channel removal. Zero-phase shift Butterworth low-pass and high-pass filters of order 2 are used. It should be noted that any filter introduces unwanted artifacts in the signal \cite{60}. So, while the GUI provides a rapid way to preliminary study a publicly available dataset, further analyses should be carried out to ensure that the result is not impacted by such artifacts. In general, one approach is to re-run all analyses after preprocessing the data with different filters, determining if the result is unaffected by that choice. Encoding and decoding (i.e., forward and backward) TRF analyses can be run with different selections of parameters (e.g., time-lag window, stimulus features). Interestingly, there is a function for limiting the number of participants and the amount of data per participant (in minutes). This functionality can be used for giving a preliminary idea of how much data is needed for the effect of interest to emerge. For ease of use, we also include a separate Stats tab with a power analysis calculator, which can be used to formally determine the minimum sample size needed to detect an effect of a given size.

\begin{figure}[bt]
\centering
\includegraphics[height=7.5cm]{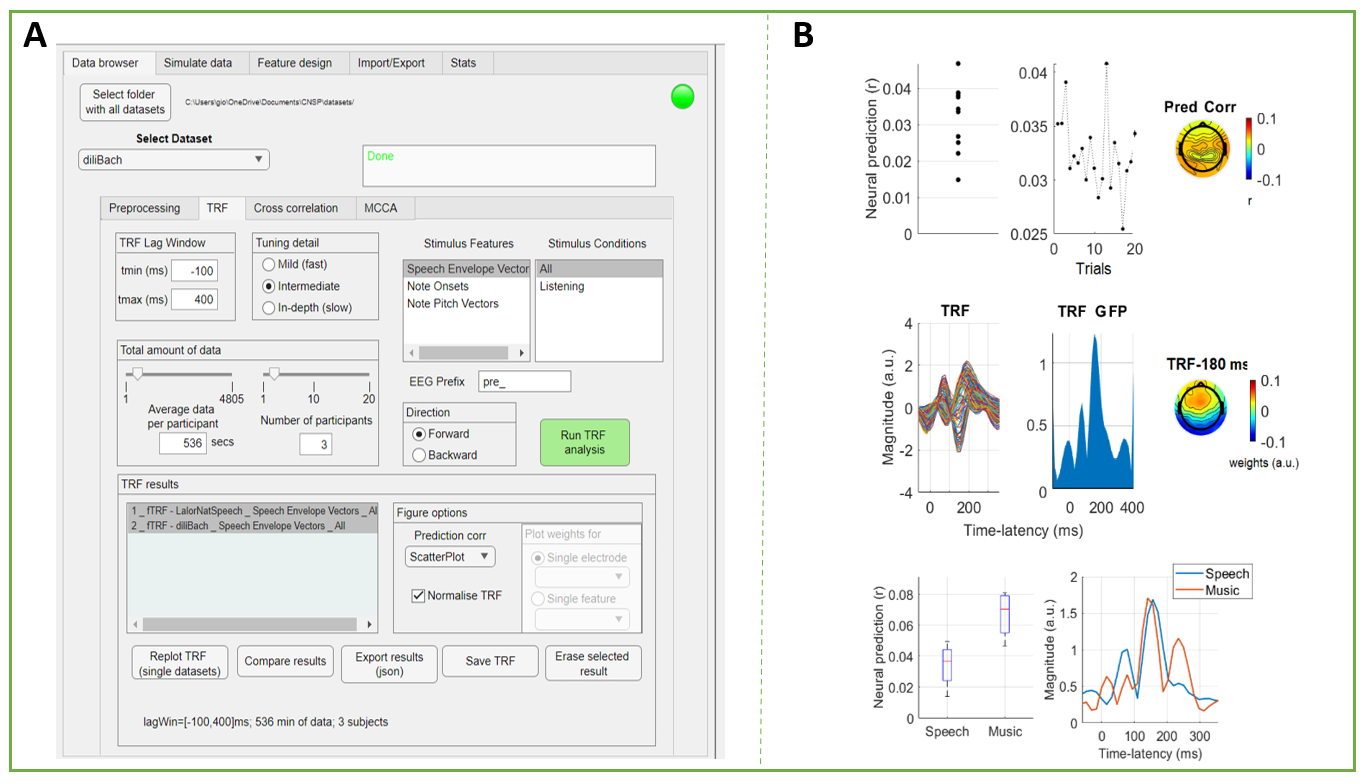}
\caption{\textbf{The Cognition and Natural Sensory Processing (CNSP) Data Browser.} (A) Screenshot of the Data Browser (left). There are five tabs coresponding to experimental analysis steps (data browser, stimulate data, feature design, import/export, and stats). Under the ‘Data Browser’ tab, the drop down menu permits dataset selection. There are four tabs corresponding to data browser settings (preprocessing, TRF, cross-correlation, MCCA). Under these settings, a user can select a specific band-pass filtering bandwidth, downsampling ratio, and TRF model hyperparameter values. There are also tabs for feature selection, data exporting, and statistical testing. (B) The GUI can visualize results across datasets or parameter configurations .The plots in the top and middle panels are the typical visualisations for forward TRF models. The top panels show, from left to right, the EEG prediction correlations for each participant (average across channels and trials), for each trial (average across channels and participants), and for each EEG channel (average across participants and trials). The middle panels show the TRF weights of a speech envelope TRF model for each channel (average across participants), the Global Field Power of the TRF weight (GFP), and the topography of the TRF weights at the peak latency of the GFP. The bottom plots compare speech and music TRFs from different datasets in terms of EEG prediction correlations and GFP.}
\end{figure}

\paragraph{Simulating}

One key issue with neural signal analysis is that the ground-truth signal is not available. In other words, we do not know what exact neural signal of interest is buried behind the large EEG/MEG noise . The only signal we have access to is the mixture of the neural signals of interest and noise, from multiple channels. Estimating the consistent neural response behind the EEG/MEG noise is the goal of methodologies such as ERPs/ERFs and TRFs, which allow us to test experimental hypotheses on that hidden neural response. However, there are scenarios where the ground-truth neural signal must first be known. One such scenario is the development of novel methodologies for neural signal analysis. For example, methodologies such as Denoising Source Separation (DSS) and Canonical Correlation Analysis (CCA) have been tested by considering various scenarios on simulated data, such as different signal-to-noise ratio (SNR), where the ground-truth signal was known by construction. Another important scenario where simulated data is necessary is the generation of numerical expectations on a given hypothesis. For example, it can be difficult to have clear expectations on what a TRF model would capture when considering certain multivariate models, especially when the stimulus features are highly correlated with each other. Therefore, simulated data can be used to examine the results when considering a specific algorithm, set of parameters, and hypothesis. 

Figure 7 depicts the graphical front-end for the simulation toolkit. The simulation consists of convolving a ground-truth TRF (designed by the user, typically based on a hypothesis or on the literature) with a given stimulus feature, and adding a specified amount of noise. The noise can either be white noise with a different magnitude, or segments of real EEG/MEG signals taken from other time-points, trials, or experiments. The approach can also be expanded to multivariate models in an additive manner. For example, EEG signals may be simulated by summing EEG noise, neural responses to the speech envelope, and neural responses to word onsets, where the neural responses are simulated by convolving each feature with their specific ground-truth TRF. The Data Browser is built on the simulation function buildSimulatedDataset.m, which can be found in the libraries folder in the CNSP GitHub. These scripts have been used recently by Chalehchaleh and colleagues \cite{62} and are conceptually similar to what was carried out in previous studies using simulation (e.g., de Cheveigné and colleagues \cite{31}).

The simulation pipeline involves three steps: 1) Design TRF (optional), 2) Select TRF, and 3) Simulate neural signal. First, a TRF can be designed from scratch by indicating a list x-y values (e.g., time and magnitude of a TRF peak). The TRF will then be obtained by fitting a curve based on those datapoints (Fig. 7A). Univariate TRFs can be combined into multivariate TRFs through the ‘combine’ function. In the ‘Select TRF’ step, the user should select a TRF (univariate or multivariate) that was either built through step 1, or which was obtained from the analysis of actual EEG/MEG data. Finally, the selected TRF (univariate or multivariate) is used to generate simulated neural data, with the addition of the selected type of noise with the given SNR. Simulated data can be generated for a single electrode or for all electrodes, for a single participant or multiple participants, depending on the specific requirements. Figure 7B shows selected examples of simulated neural traces based on different combinations of parameters.

\begin{figure}[bt]
\centering
\includegraphics[height=7cm]{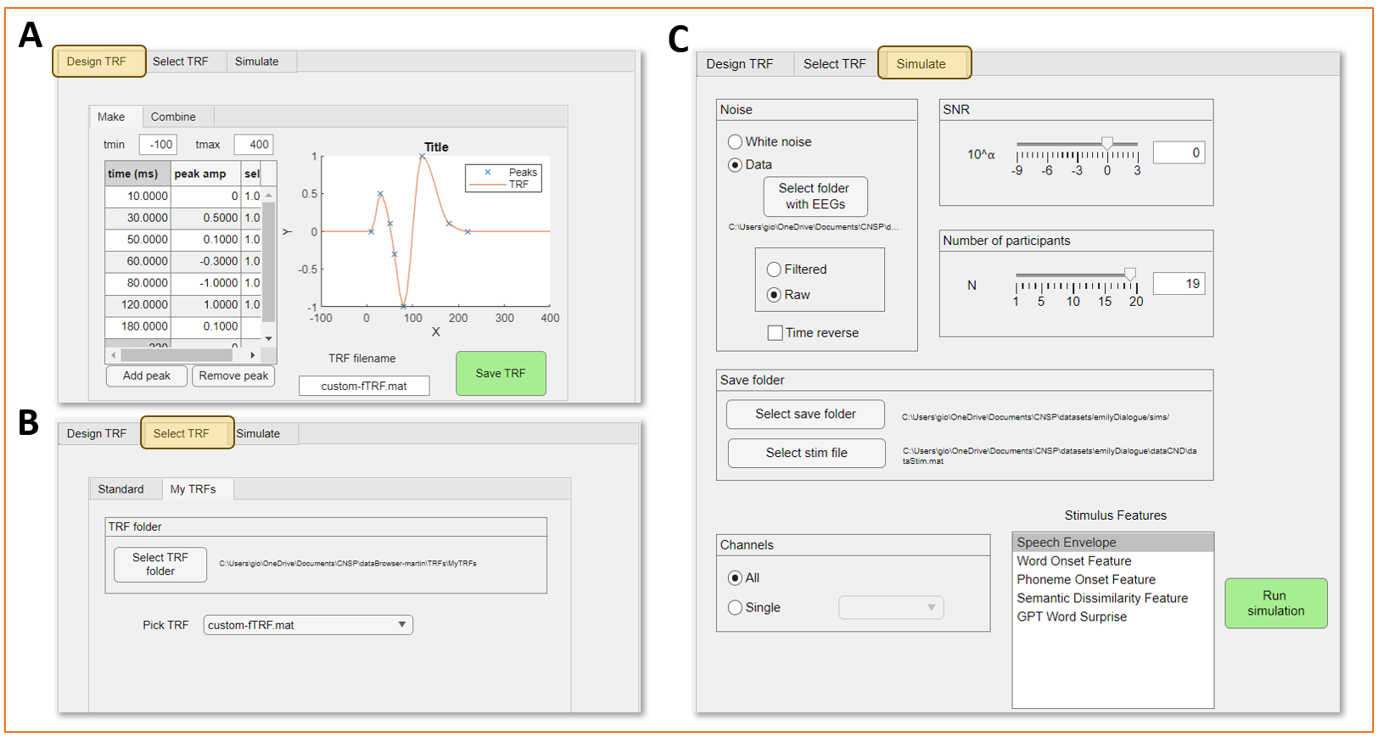}
\caption{\textbf{Simulation toolkit.} The simulation front-end is organised into three tabs. (A) First, the user can design a new temporal response function (TRF) by interpolating a list of given datapoints. A typical speech envelope TRF is designed by indicating datapoints describing the P1, N1, and P2 components. Datapoints should capture the magnitude and latency of each component as well as the time needed for the component to return to baseline. Multivariate TRFs can also be built by concatenating pre-built TRF. (B) The user may decide to utilise a TRF they built with the TRF designer, or one of the standard TRFs that are included with the simulation toolkit. (C) The user may then proceed to the ‘Simulate’ tab. The selected TRF will be convolved with an existing stimulus feature time-series (within dataStim.mat), producing a ground-truth EEG/MEG trace. Noise will be added with the selected signal-to-noise-ratio (SNR) to simulate more realistic EEG/MEG data. The noise can be white noise or real EEG/MEG data, preferably from another experiment. The resulting simulated data can then be analysed with the data browser as a dataset.}
\end{figure}

\paragraph{Web-based Data Browser}
The work presented above primarily relies on MATLAB software. While the data browser GUI eliminates the client-side need for a MATLAB license, the user remains in charge of appropriately installing the application and downloading the datasets. However, the rapidly growing number of publicly available datasets will likely challenge the use of a GUI in local due to space limitations. Here we present the first web-based data browser for continuous sensory neurophysiology. While the functionalities will be similar to those of the MATLAB-based GUI, the current version focuses on forward and backward TRF models only. This proof-of-concept web-application stores datasets and runs analyses directly in the cloud, meaning that the user is no longer limited to the computational constraints of their own machine. In turn, the use of Python language for implementing the web-based data browser, relying on the Python implementation of the mTRF-Toolbox \cite{30,49}, enhances the accessibility of this project. The web-based CNSP Data Browser can be accessed from the dedicated page on the documentation https://cnsp-resources.readthedocs.io/en/latest/
databrowserPage.html.

\begin{figure}[bt]
\centering
\includegraphics[height=10cm]{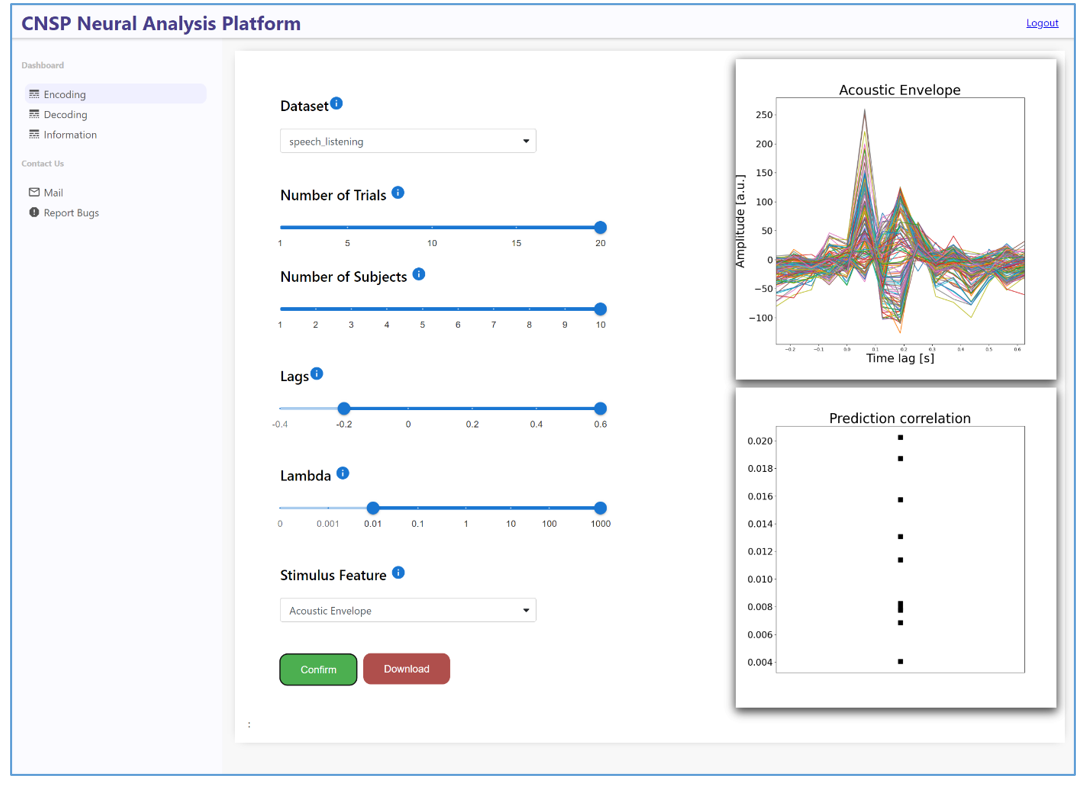}
\caption{\textbf{The Web-based CNSP Data Browser.} The web-based data browser aims to serve the same functions of the MATLAB GUI, but from an Internet browser. The web-based data browsed, which is Python-based, currently supports mTRF forward and backward models i.e., encoding and decoding models. One of the key advantages of this innovative piece of software is that it enables the use of TRF analyses without the need for local installation or download; expanding accessiblity to continous-event neural data. Future browser iterations will add functionalities from other toolkits, such as CCA, NAPLib, and Eelbrain, enabling the use of a large set of methodologies on the same dataset. }
\end{figure}

\paragraph{Engagement with the research community}
The concepts and resources presented here are the result of thorough discussions and reflections carried out at the CNSP workshops in 2021, 2022, and 2023. Such discussions involved, at different levels, the 322 participants of the workshops, 19 speakers, and 6 organisers. The 322 participants their university or company affiliation to be in the EU (37\%), 34\% USA, 8\% UK, and with some representation (<5\%) of Canada, Australia, China, Iran, Uruguay, Greece, South Korea, Japan, Israel, India, and Switzerland. Most participants were PhD students (48\%), postdoctoral researchers (22\%), and faculty members (9\%), with the remainder consisting of undergraduate and master’s students, research assistants, and industry researchers. Finally, most participants indicated that they use a combination of multiple programming/scripting languages for their work, with the majority of them using MATLAB (80\%), Python (54\%), and R (45\%). These numbers are indicative of a wide interest in the topics covered in the CNSP-Workshops, with a diverse representation of the research community. While we think that this constitutes a solid starting point for the application and further development of the framework described in this manuscript, there are indeed several areas of improvement that we will consider. First of all, further representation of researchers in Asia would likely be possible be adjusting the time the virtual CNSP workshop is carried out, which was previously not optimal for countries in east Asia. Further representation of other parts of the world (e.g., south America) will be sought by expanding the mailing lists that are targeted when advertising the workshop. Note that, while scientists from anywhere have access to the resources already, it is our goal to continue engaging with the participants directly through the CNSP workshops, encouraging a more balanced exchange of ideas and opinions on the way forward.

\section{Discussion}

Here, we provide a comprehensive set of resources for the re-analysis and sharing of neural data involving continuous sensory stimuli. The guidelines in this manuscript encompass the entire analytic pipeline, from data storage, to data analysis, import/export functions connecting the present work with other resources in the literature, and simulation functions for the formulation of specific hypotheses. The guidelines in this manuscript are complemented with educational resources, such as tutorial scripts and video-tutorials, which are available on https://cnspinitiative.net.

The proposed approach is in line with the FAIR principles (Findable, Accessible, Interoperable and Reusable) \cite{63}. The set of resources and guidelines are built around the principle of reusability. Interoperability is ensured in multiple ways. The proposed CND structure and conversion mechanisms from/to BIDS ensure that the data can be easily analysed with existing tools across various operating systems and programming languages. The CNSP Data Browser provides executables for various operating systems, bypassing the need for a MATLAB licence. Finally, the web-based data browser is a proof-of-concept front-end for combining and comparing tools from different libraries and programming languages, further contributing to interoperability. Regarding the ‘finding’ and ‘accessible’ principle, while addressing these principles directly was beyond the scope of the present article, we encourage the adherence to these principles, for example by relying on reliable and findable repositories providing unique identifiers, such as OpenNeuro (https://openneuro.org/) and Dryad (https://datadryad.org/). 

The key intention behind this article is to simplify analysis and data sharing and re-use by establishing an appropriate pipeline and sharing tools that reduce the amount of work required for new experiments. While working with custom methods and data structure might be quicker at producing results initially, a standardised pipeline is of great advantage at the time of data sharing, when re-analysing existing data, or even when the user intends to compare results on different datasets or to use resources across different toolboxes on the same data, while also reducing the risk for errors. The proposed approach makes it easy to try different methodologies and parameter choices on the same dataset which heavily simplifies the direct comparison of methodologies from different teams. Functionalities such as the simulation toolkit in the GUI aim to encourage the use of simulations. For example, studies with multivariate TRF could greatly benefit from numerical simulations. However, one important observation is that offering easily accessible analytic functionalities through a GUI carries risks. For example, a user might use the GUI as a black-box, without truly understanding what it is doing and, thus, potentially leading to misinterpretation of the results. The GUI design attempts to mitigate that risk, by putting an emphasis on speeding-up the analysis pipeline, but without providing the full range of functionalities in the actual CNSP libraries. So, the present version of the data browser offers the necessary tools for running typical TRF (and other) analyses, replicating existing results, and running simulations rapidly. But all of that is constrained so that only the typical analyses can be carried out, thus minimising the risk of misuse. In that sense, the educational resources such as the tutorials, video-tutorials, and workshop are key as they empower the users, giving them the tools for fully availing of the CNSP resources. 

The cohesive set of resources presented in this article highlights the importance of truly committing to the ‘reusability’ FAIR principle. This is unfortunately not always the case, as the data shared in this research domain is often limited to the data that is necessary for replicating the figures, but only by sharing processed data or even just the result of the analyses. For example, an EEG study involving a speech listening task should ideally be shared by including the original audio files and the raw EEG data when possible, so that it can be re-analysed in various different ways. Instead, some existing datasets (including early datasets from the authors of this manuscript) did not include the raw audio files, for example due to copyright limitations. That and other similar issues should be carefully taken into consideration at the experimental design stage, now that the possibility and benefits of data reusability are established. As a result, we encourage journals and reviewers to take reusability more into consideration in the future. In other words, sharing a dataset in a standardised format like BIDS should not be taken as a guarantee in itself that the data is shared appropriately. Instead, the data sharing statement should include considerations on the possible reusability of the dataset, thus reflecting on the best way of sharing data beyond replicability. Another important issue is that the code used for the analyses is often not fully shared during the journal publication process. Instead, we encourage journals to be more strict on that point, which is essential for result replication i.e., scientific papers should include either the custom code used or the exact reference to the tools used, such as a DOI, version number, or at least the commit index (e.g., GitHub commit index).

We expect that future research in this domain will lead to a large set of data that will open new possibilities for both theoretical investigations and methodological development, benefiting neurophysiology research at large. In these re-analysis scenarios, it is important to stress the key role of result replication. In fact, apparently valid results may emerge by chance due to multiple re-analyses of the same dataset from different individuals. For this reason, it is extremely important to replicate the results on another publicly available dataset or, even better, on a new dataset.

\paragraph{Limitations and future work}

The present work represents the first comprehensive and cohesive set of analytic resources, from data standardisation to data analysis and simulation, that is specific to continuous sensory neural data. In doing so, the main contribution of this work is the attempt to build resources and guidelines that are tailored to the specific needs of this field of research. Indeed, the resources provided are constantly improved as the field advances, making this specific manuscript not immune to technical limitations. Here, we discuss these limitations with the goal of pin-pointing important directions for future work in this area. 

One limitation with this work is that the majority of the code and resources shared for neural data analysis rely on the MATLAB language, which limits access to these resources to individuals with prior expertise with that language, and that hold a MATLAB license. That issue was mitigated by producing the executable graphical interface, which was built with MATLAB but can run as a stand-alone piece of software. Furthermore, we also developed a Python-based web-based data browser. While the web-based data browser only has limited functionalities at the time this was written, future work will substantially extend its functions by giving it full access to a variety of relevant toolboxes, such as the mTRF-Toolbox, Eelbrain-toolbox, NAPLib, and others. In doing so, this tool aims to encapsulate the open-science change in sensory neuroscience, implementing the relevant FAIR principles and opening new possibilities beyond result replication and re-analysis, such as developments involving the availability of big, aggregated datasets.

The description in this manuscript focuses on speech and music listening datasets. Future research should test the validity of the guidelines and resources onto other relevant domains in sensory neuroscience, starting from research involving rapid continuous stimuli in other sensory modalities, such as visual and tactile modalities \cite{64}. The work could also be expanded to other domains involving motor movements, where methodologies such as TRFs and CCA can be used \cite{65}. As future work will continue developing these tools by incorporating other methodologies and additional datasets, the metadata will be expanded to keep track of the version number of the tools used for storing and manipulating the data. 

Beyond the specific resources, scripts, and data structures, we contend that the most important and long-lasting overarching idea will be facilitating the use of lab specific resources (scripts and data) while connecting resources from different teams, so that they are interoperable. Important lessons can be learnt from other research fields, such as bioinformatics, guiding future development in sensory neuroscience toward more efficient research framework, encouraging the use of these domain-specific guidelines for data sharing while avoiding risks such as excessively rigid data sharing policies, which could do more harm than good. Indeed, a collective effort is necessary to achieve that goal, and we invite the research community to contribute to this endeavour by sharing data, well-documented scripts, and tutorials, enriching the CNSP resources with their methodologies and data.

\section{Methods}

\paragraph{Continuous-event Neural Data (CND)}

A main folder structure is provided containing all the experimental files, organised into ‘code’, ‘tutorials’, ‘datasets’, and ‘libs’ (libraries) folders. For simplicity, we advise users to have one main folder for each study. If multiple experiments are being considered, such as when using the Data Browser, it is also possible to include many datasets in the ‘datasets’ folder. Each dataset includes a ‘dataCND’ folder, containing as many ‘dataStimX.mat’ (stimulus files) and ‘dataSubX.mat’ (neural data) files as the number of participants. If all participants were exposed to identical stimuli, then a single ‘dataStim.mat’ file can be used. In that case, stimuli might have been presented in different orders for distinct participants (e.g., [1,2,3,…], [2,3,1,…]) and, as such, the neural data segments in ‘dataSubX.mat’ will have to be sorted to match the order of the single stimulus file. The ‘rawStimuli’ and ‘rawNeural’ data folders contain the unstructured original data and custom processing code (e.g., raw2cnd, feature extraction in Fig. 2A). Both folders are optional but recommended. For example, in the case of relatively large datasets (>10 GB), we suggest experimenters share two versions of the data, with and without the original raw files. A how-to guide is provided on converting raw data into CND (see bids2cnd function). 

\paragraph{Storing raw data in CND format: practical considerations}

Here, we discuss a bdf2cnd conversion for a scenario where EEG signals were recorded during speech listening, with audio segments presented in a random order (different order for distinct participants). The synchronisation protocol is simple, as it involves only one type of trigger indicating the start of an audio segment (as in Figure 3). The code of that trigger corresponds to the index of the audio file (e.g., audio1.wav). To save a new dataset into CND, the first step is to create a folder structure as in Figure 5A. We strongly recommend the naming of files as ‘audio1.wav’, ‘audio2.wav’, and so on, while also including the original source of those stimuli or corresponding conditions in the documentation, but not in the filename. The simplest way to build a CND dataset is to create a ‘dataStim’ and ‘dataSub’ file for each participant. ‘dataSub’ contains a structure with the epoched neural data and meta-data, such as the sampling frequency (‘fs’) and the channel location information (‘chanlocs’; see Figure 5B). The epoching simply consists of chunking the neural data into segments, where each segment paired and synchronised with a specific audio input. This data matrix will correspond to the stimulus data matrix in ‘dataStim’, which will have the same number of trials and start samples. Note that the CND files can contain raw epoched EEG data, which may be at a different sampling rate than the stimuli. That will necessarily be corrected at the preprocessing stage, where it is ensured that the same sampling rate and number of samples are present in the stimulus and neural data. Stimulus features are stored as data vectors (when univariate) or data matrices (when multivariate), such as the sound envelope (timeSamples x 1) and spectrogram (timeSamples x numberOfBands) respectively. Rather than adopting a sparse representation, CND requires a vector/matrix representation preserving all datapoints, including the zeroes, which is necessary for toolboxes such as the mTRF-Toolbox and NoiseTools. A full list of specifications is available on https://cnsp-resources.readthedocs.io, with example scripts and video-tutorials.

\paragraph{The web-based CNSP data browser}

The web-based CNSP data browser combines a frontend written in React, a popular user-interface JavaScript library, and a backend in Flask, a micro web framework for Python. The frontend employs React's component-based structure for a dynamic user interface, while Flask handles running the analyses and communication. Google App Engine ensures seamless deployment, auto-scaling, and load balancing while Google Cloud Storage securely manages the stored datasets and images from previous analyses, reducing the computational load. This integrated approach offers a scalable, responsive, and reliable web application architecture that holds promise for future expansion and improvement. 

\paragraph{Code availability}

All code is available on GitHub (https://github.com/CNSP-Workshop) and on the website of the CNSP initiative (https:// cnspinitiative.net). The specific version of the code at the time of publication of this manuscript was shared via FigShare (https://figshare.com/s/4cb6d497ddde5fd0a29f). 

\paragraph{Data availability}

This work used publicly available data. That data, which were stored in various custom data structures, were converted into the proposed CND data structured and re-shared on the website of the CNSP initiative (https://cnspinitiative.net). As that website is updated on a continuous basis, we also re-shared selected datasets co-authored by GDL and EL here https://figshare.com/s/4cb6d497ddde5fd0a29f. Please refer to the original license and citation of each dataset when using them.

\paragraph{Author contributions}

GDL and AN wrote the first draft of the manuscript. MC, NZ, GC, MW, AI, SH, GB, and EL edited the manuscript. GDL, AN, NZ, and GC revised the manuscript. GDL, AN, MJC, NZ, SH, GC organised the CNSP workshops, producing tutorials and video-tutorials. MW and GDL built the GUI and the simulation toolkit. RM and AI built the web-based data browser. GDL, AN, MC, and SC contributed to the CND data structure. SC, MW, and GDL connected BIDS and CND. GDL and AI wrote the online documentation. GDL and AI prepared the figures in this manuscript. GB wrote Box 1, discussing the key role of open science in bioinformatics.

\paragraph{Competing Interests}

The authors do not have competing interests to report.

\paragraph{Funding sources}

This research was supported by the Science Foundation Ireland under Grant Agreement No. 13/RC/2106\_P2 at the ADAPT SFI Research Centre at Trinity College Dublin. ADAPT, the SFI Research Centre for AI-Driven Digital Content Technology, is funded by Science Foundation Ireland through the SFI Research Centres Programme. SH was supported by a National Institute of Health (NIH) T32 Trainee Grant No. 5T32DC000038-27 and the National Science Foundation (NSF) Graduate Research Fellowship Program under Grant No. DGE1745303. GC was supported by an Advanced European Research Council grant (NEUME, 787836) 

\paragraph{Acknowledgements}

The authors thank Gavin Moore Mischler and Christian Brodbeck for writing CND loading functions for their respective toolboxes, NAPLib and Eelbrain.

\printendnotes

\bibliography{main}

\begin{thebibliography}{63}
\providecommand{\natexlab}[1]{#1}
\providecommand{\url}[1]{\texttt{#1}}
\providecommand{\urlprefix}{}

\bibitem[{Clark(2016)Clark, A.}]{1}
Clark A.
\newblock Surfing Uncertainty: Prediction, Action, and the Embodied Mind.
\newblock Oxford University Press; 2016.
\newblock \urlprefix\url{https://books.google.ie/books?id=Yoh2CgAAQBAJ}.

\bibitem[{Kutas and Federmeier(2011)Kutas, Marta and Federmeier, Kara D.}]{2}
Kutas M, Federmeier KD.
\newblock Thirty years and counting: Finding meaning in the N400 component of the event-related brain potential (ERP).
\newblock Annual Review of Psychology 2011;62:621--647.
\newblock \urlprefix\url{http://www.ncbi.nlm.nih.gov/pubmed/20809790 http://www.pubmedcentral.nih.gov/articlerender.fcgi?artid=PMC4052444}.

\bibitem[{Paller et~al.(1992)Paller, K. A. and McCarthy, G. and Wood, C. C.}]{3}
Paller KA, McCarthy G, Wood CC.
\newblock Event-related potentials elicited by deviant endings to melodies.
\newblock Psychophysiology 1992;29(2):202--6.
\newblock \urlprefix\url{http://www.ncbi.nlm.nih.gov/pubmed/1635962}.

\bibitem[{Yabe et~al.(1997)Yabe, Hirooki and Tervaniemi, Mari and Reinikainen, Kalevi and N{\"a}{\"a}t{\"a}nen, Risto}]{4}
Yabe H, Tervaniemi M, Reinikainen K, N{\"a}{\"a}t{\"a}nen R.
\newblock Temporal window of integration revealed by MMN to sound omission.
\newblock NeuroReport 1997;8(8):1971--1974.
\newblock \urlprefix\url{https://pubmed.ncbi.nlm.nih.gov/9223087/}.

\bibitem[{Theunissen et~al.(2001)Theunissen, Fr{'e}d{'e}ric E. and David, Stephen V. and Singh, Nandini C. and Hsu, Anne and Vinje, William E. and Gallant, Jack L.}]{5}
Theunissen FE, David SV, Singh NC, Hsu A, Vinje WE, Gallant JL.
\newblock Estimating spatio-temporal receptive fields of auditory and visual neurons from their responses to natural stimuli.
\newblock Network: Computation in Neural Systems 2001;12(3):289--316.

\bibitem[{David et~al.(2007)David, Stephen V. and Mesgarani, Nima and Shamma, Shihab A.}]{6}
David SV, Mesgarani N, Shamma SA.
\newblock Estimating sparse spectro-temporal receptive fields with natural stimuli.
\newblock Network: Computation in Neural Systems 2007;18(3):191--212.
\newblock \urlprefix\url{https://www.tandfonline.com/doi/full/10.1080/09548980701609235}.

\bibitem[{Bonte et~al.(2006)Bonte, Milene and Parviainen, Tiina and Hyt{\"o}nen, Kaisa and Salmelin, Riitta}]{7}
Bonte M, Parviainen T, Hyt{\"o}nen K, Salmelin R.
\newblock Time course of top-down and bottom-up influences on syllable processing in the auditory cortex.
\newblock Cerebral Cortex 2006;16(1):115--123.
\newblock \urlprefix\url{http://cercor.oxfordjournals.org/content/16/1/115.full.pdf}.

\bibitem[{Gold et~al.(2019)Gold, Benjamin P. and Pearce, Marcus T. and Mas-Herrero, Ernest and Dagher, Alain and Zatorre, Robert J.}]{8}
Gold BP, Pearce MT, Mas-Herrero E, Dagher A, Zatorre RJ.
\newblock Predictability and uncertainty in the pleasure of music: a reward for learning?
\newblock The Journal of Neuroscience 2019;p. 0428--19.

\bibitem[{Fahmie et~al.(2023)Fahmie, Tara A. and Rodriguez, Nicole M. and Luczynski, Kevin C. and Rahaman, Javid A. and Charles, Brinea M. and Zangrillo, Amanda N.}]{9}
Fahmie TA, Rodriguez NM, Luczynski KC, Rahaman JA, Charles BM, Zangrillo AN.
\newblock Toward an explicit technology of ecological validity.
\newblock Journal of Applied Behavior Analysis 2023;56(2):302--322.
\newblock \urlprefix\url{https://onlinelibrary.wiley.com/doi/abs/10.1002/jaba.972}.

\bibitem[{Tervaniemi(2023)Tervaniemi, Mari}]{10}
Tervaniemi M.
\newblock The neuroscience of music; towards ecological validity.
\newblock Trends in Neurosciences 2023;46(5):355--364.
\newblock \urlprefix\url{https://doi.org/10.1016/j.tins.2023.03.001}.

\bibitem[{Broderick et~al.(2018)Broderick, MP and Anderson, AJ and Di Liberto, GM and Crosse, MJ and Lalor, EC}]{11}
Broderick M, Anderson A, Di~Liberto G, Crosse M, Lalor E, Data from: electrophysiological correlates of semantic dissimilarity reflect the comprehension of natural, narrative speech. Dryad Digital Repository. Published online February 23, 2018; 2018.

\bibitem[{Di~Liberto et~al.(2020)Di Liberto, Giovanni M. and Pelofi, Claire and Bianco, Roberta and Patel, Prachi and Menhta, Ashesh D. and Herrero, Jose L. and Shamma, Shihab A. and Mesgarani, Nima and Mehta, Ashesh D. and Herrero, Jose L. and de Cheveign{'e}, Alain and Shamma, Shihab A. and Mesgarani, Nima}]{12}
Di~Liberto GM, Pelofi C, Bianco R, Patel P, Menhta AD, Herrero JL, et~al.
\newblock Cortical encoding of melodic expectations in human temporal cortex.
\newblock eLife 2020;9.

\bibitem[{Hale et~al.(2018)Hale, John and Dyer, Chris and Kuncoro, Adhiguna and Brennan, Jonathan R.}]{13}
Hale J, Dyer C, Kuncoro A, Brennan JR.
\newblock Finding syntax in human encephalography with beam search.
\newblock In: ACL 2018 - 56th Annual Meeting of the Association for Computational Linguistics, Proceedings of the Conference (Long Papers), vol.~1 Association for Computational Linguistics; 2018. p. 2727--2736.

\bibitem[{Das et~al.(2016)Das, Neetha and Biesmans, Wouter and Bertrand, Alexander and Francart, Tom}]{14}
Das N, Biesmans W, Bertrand A, Francart T.
\newblock The effect of head-related filtering and ear-specific decoding bias on auditory attention detection.
\newblock Journal of Neural Engineering 2016;13(5):056014--056014.
\newblock \urlprefix\url{http://www.ncbi.nlm.nih.gov/pubmed/27618842 http://stacks.iop.org/1741-2552/13/i=5/a=056014?key=crossref.96b84e130f6e736c4b8510973a6be747}.

\bibitem[{Di~Liberto et~al.(2015)Di Liberto, G. M. and O'Sullivan, J. A. and Lalor, E. C.}]{15}
Di~Liberto GM, O'Sullivan JA, Lalor EC.
\newblock Low-frequency cortical entrainment to speech reflects phoneme-level processing.
\newblock Current Biology 2015;25(19).

\bibitem[{Ding and Simon(2014)Ding, Nai and Simon, Jonathan Z.}]{16}
Ding N, Simon JZ.
\newblock Cortical Entrainment to Continuous Speech: Functional Roles and Interpretations.
\newblock Frontiers in human neuroscience 2014;8.
\newblock \urlprefix\url{http://www.frontiersin.org/Journal/Abstract.aspx?s=537&name=human_neuroscience&ART_DOI=10.3389/fnhum.2014.00311 http://journal.frontiersin.org/article/10.3389/fnhum.2014.00311/pdf}.

\bibitem[{Kern et~al.(2022)Kern, Pius and Heilbron, Micha and de Lange, Floris P. and Spaak, Eelke}]{17}
Kern P, Heilbron M, de~Lange FP, Spaak E.
\newblock Cortical activity during naturalistic music listening reflects short-range predictions based on long-term experience.
\newblock eLife 2022;11:e80935.
\newblock \urlprefix\url{https://doi.org/10.7554/eLife.80935}.

\bibitem[{Jessica~Tan et~al.(2022)Jessica Tan, S. H. and Kalashnikova, Marina and Di Liberto, Giovanni M. and Crosse, Michael J. and Burnham, Denis}]{18}
Jessica~Tan SH, Kalashnikova M, Di~Liberto GM, Crosse MJ, Burnham D.
\newblock Seeing a talking face matters: The relationship between cortical tracking of continuous auditory‐visual speech and gaze behaviour in infants, children and adults.
\newblock NeuroImage 2022;256:119217.
\newblock \urlprefix\url{https://www.sciencedirect.com/science/article/pii/S105381192200341X}.

\bibitem[{Crosse et~al.(2016)Crosse, M. J. and Di Liberto, G. M. and Lalor, E. C.}]{19}
Crosse MJ, Di~Liberto GM, Lalor EC.
\newblock Eye can hear clearly now: Inverse effectiveness in natural audiovisual speech processing relies on long-term crossmodal temporal integration.
\newblock Journal of Neuroscience 2016;36(38).

\bibitem[{Lalor et~al.(2006)Lalor, Edmund C. and Pearlmutter, Barak A. and Reilly, Richard B. and McDarby, Gary and Foxe, John J.}]{20}
Lalor EC, Pearlmutter BA, Reilly RB, McDarby G, Foxe JJ.
\newblock The VESPA: a method for the rapid estimation of a visual evoked potential.
\newblock NeuroImage 2006;32(4):1549--1561.
\newblock \urlprefix\url{http://ac.els-cdn.com/S1053811906006434/1-s2.0-S1053811906006434-main.pdf?_tid=ff77a230-a642-11e4-8c11-00000aacb35e&acdnat=1422376921_3d415c577602776e8b0c6a9e8425ac29}.

\bibitem[{Guilleminot and Reichenbach(2022)Guilleminot, Pierre and Reichenbach, Tobias}]{21}
Guilleminot P, Reichenbach T.
\newblock Enhancement of speech-in-noise comprehension through vibrotactile stimulation at the syllabic rate.
\newblock Proceedings of the National Academy of Sciences 2022;119(13):e2117000119.

\bibitem[{Kraemer et~al.(2005)Kraemer, David J. M. and Macrae, C. Neil and Green, Adam E. and Kelley, William M.}]{22}
Kraemer DJM, Macrae CN, Green AE, Kelley WM.
\newblock Sound of silence activates auditory cortex.
\newblock Nature 2005;434(7030):158--158.
\newblock \urlprefix\url{https://www.nature.com/articles/434158a}.

\bibitem[{Di~Liberto et~al.(2021)Di Liberto, Giovanni M. and Marion, Guilhem and Shamma, Shihab A.}]{23}
Di~Liberto GM, Marion G, Shamma SA.
\newblock The Music of Silence: Part II: Music Listening Induces Imagery Responses.
\newblock The Journal of Neuroscience 2021;41(35):7449.
\newblock \urlprefix\url{http://www.jneurosci.org/content/41/35/7449.abstract}.

\bibitem[{Joutsiniemi and Hari(1989)Joutsiniemi, S. L and Hari, R.}]{24}
Joutsiniemi SL, Hari R.
\newblock Omissions of Auditory Stimuli May Activate Frontal Cortex.
\newblock European Journal of Neuroscience 1989;1(5):524--528.
\newblock \urlprefix\url{https://pubmed.ncbi.nlm.nih.gov/12106138/}.

\bibitem[{Heilbron and Chait(2018)Heilbron, Micha and Chait, Maria}]{25}
Heilbron M, Chait M, Great Expectations: Is there Evidence for Predictive Coding in Auditory Cortex?
\newblock Elsevier Ltd; 2018.
\newblock \urlprefix\url{https://pubmed.ncbi.nlm.nih.gov/28782642/}.

\bibitem[{Khalighinejad et~al.(2017)Khalighinejad, Bahar and Cruzatto da Silva, Guilherme and Mesgarani, Nima}]{26}
Khalighinejad B, Cruzatto~da Silva G, Mesgarani N.
\newblock Dynamic Encoding of Acoustic Features in Neural Responses to Continuous Speech.
\newblock The Journal of Neuroscience 2017;.

\bibitem[{Broderick et~al.(2018)Broderick, M. P. and Anderson, A. J. and Di Liberto, G. M. and Crosse, M. J. and Lalor, E. C.}]{27}
Broderick MP, Anderson AJ, Di~Liberto GM, Crosse MJ, Lalor EC.
\newblock Electrophysiological Correlates of Semantic Dissimilarity Reflect the Comprehension of Natural, Narrative Speech.
\newblock Current Biology 2018;.

\bibitem[{Brodbeck et~al.(2018)Brodbeck, Christian and Hong, L. Elliot and Simon, Jonathan Z.}]{28}
Brodbeck C, Hong LE, Simon JZ.
\newblock Rapid Transformation from Auditory to Linguistic Representations of Continuous Speech.
\newblock Current Biology 2018;28(24):3976--3983.e5.
\newblock \urlprefix\url{https://www.cell.com/current-biology/fulltext/S0960-9822(18)31409-X?_returnURL=https%3A%2F%2Flinkinghub.elsevier.com%2Fretrieve%2Fpii%2FS096098221831409X%3Fshowall%3Dtrue https://linkinghub.elsevier.com/retrieve/pii/S096098221831409X}.

\bibitem[{Heilbron et~al.(2022)Heilbron, Micha and Armeni, Kristijan and Schoffelen, Jan-Mathijs and Hagoort, Peter and de Lange, Floris P.}]{29}
Heilbron M, Armeni K, Schoffelen JM, Hagoort P, de~Lange FP.
\newblock A hierarchy of linguistic predictions during natural language comprehension.
\newblock Proceedings of the National Academy of Sciences 2022;119(32):e2201968119.
\newblock \urlprefix\url{https://www.pnas.org/doi/abs/10.1073/pnas.2201968119}.

\bibitem[{Crosse et~al.(2016)Crosse, M. J. and Di Liberto, G. M. and Bednar, A. and Lalor, E. C.}]{30}
Crosse MJ, Di~Liberto GM, Bednar A, Lalor EC.
\newblock The multivariate temporal response function (mTRF) toolbox: A MATLAB toolbox for relating neural signals to continuous stimuli.
\newblock Frontiers in Human Neuroscience 2016;10(NOV2016).

\bibitem[{de~Cheveign{'e} et~al.(2018)de Cheveign{'e}, Alain and Wong, Daniel E. and Di Liberto, Giovanni M. and Hjortkjær, Jens and Slaney, Malcolm and Lalor, Edmund}]{31}
de~Cheveign{'e} A, Wong DE, Di~Liberto GM, Hjortkjær J, Slaney M, Lalor E.
\newblock Decoding the auditory brain with canonical component analysis.
\newblock NeuroImage 2018;172:206--216.
\newblock \urlprefix\url{https://www.sciencedirect.com/science/article/pii/S1053811918300338}.

\bibitem[{Shannon et~al.(1995)Shannon, R. V. and Zeng, F. G. and Kamath, V. and Wygonski, J. and Ekelid, M.}]{32}
Shannon RV, Zeng FG, Kamath V, Wygonski J, Ekelid M.
\newblock Speech recognition with primarily temporal cues.
\newblock Science 1995;270(5234):303--304.

\bibitem[{Rosen(1992)Rosen, S.}]{33}
Rosen S.
\newblock Temporal information in speech: acoustic, auditory and linguistic aspects.
\newblock Philos Trans R Soc Lond B Biol Sci 1992;336(1278):367--73.

\bibitem[{Crosse et~al.(2021)Crosse, Michael J and Zuk, Nathaniel J and Di Liberto, Giovanni M and Nidiffer, Aaron R and Molholm, Sophie and Lalor, Edmund C}]{35}
Crosse MJ, Zuk NJ, Di~Liberto GM, Nidiffer AR, Molholm S, Lalor EC.
\newblock Linear modeling of neurophysiological responses to speech and other continuous stimuli: methodological considerations for applied research.
\newblock Frontiers in neuroscience 2021;15:705621.

\bibitem[{Obleser and Kayser(2019)Obleser, Jonas and Kayser, Christoph}]{36}
Obleser J, Kayser C.
\newblock Neural entrainment and attentional selection in the listening brain.
\newblock Trends in cognitive sciences 2019;23(11):913--926.

\bibitem[{Caucheteux et~al.(2022)Caucheteux, Charlotte and Gramfort, Alexandre and King, Jean-R{'e}mi}]{37}
Caucheteux C, Gramfort A, King JR.
\newblock Deep language algorithms predict semantic comprehension from brain activity.
\newblock Scientific Reports 2022;12(1):16327.
\newblock \urlprefix\url{https://doi.org/10.1038/s41598-022-20460-9}.

\bibitem[{Lalor et~al.(2009)Lalor, E. C. and Power, A. J. and Reilly, R. B. and Foxe, J. J.}]{38}
Lalor EC, Power AJ, Reilly RB, Foxe JJ.
\newblock Resolving Precise Temporal Processing Properties of the Auditory System Using Continuous Stimuli.
\newblock Journal of Neurophysiology 2009;102(1):349--359.
\newblock \urlprefix\url{http://jn.physiology.org/content/jn/102/1/349.full.pdf}.

\bibitem[{O'Sullivan et~al.(2014)O'Sullivan, James A. and Power, Alan J. and Mesgarani, Nima and Rajaram, Siddharth and Foxe, John J. and Shinn-Cunningham, Barbara G. and Slaney, Malcolm and Shamma, Shihab A. and Lalor, Edmund C.}]{39}
O'Sullivan JA, Power AJ, Mesgarani N, Rajaram S, Foxe JJ, Shinn-Cunningham BG, et~al.
\newblock Attentional Selection in a Cocktail Party Environment Can Be Decoded from Single-Trial EEG.
\newblock Cerebral Cortex 2014;p. bht355--bht355.

\bibitem[{Di~Liberto et~al.(2021)Di Liberto, Giovanni M. and Pelofi, Claire and Bianco, Roberta and Patel, Prachi and Mehta, Ashesh D. and Herrero, Jose L. and De Cheveign{'e}, Alain and Shamma, Shihab A. and Mesgarani, Nima}]{40}
Di~Liberto GM, Pelofi C, Bianco R, Patel P, Mehta AD, Herrero JL, et~al., Cortical encoding of melodic expectations in human temporal cortex; 2021.

\bibitem[{Di~Liberto et~al.(2023)Di Liberto, G. M. and Goswami, U. and Attaheri, A. and Choisdealbha {'A}, N. and Rocha, S. and Mead, N. and Olawole-Scott, H. and Grey, C.}]{41}
Di~Liberto GM, Goswami U, Attaheri A, Choisdealbha~{'A} N, Rocha S, Mead N, et~al., Data and code from “Emergence of the cortical encoding of phonetic features in the first year of life".
\newblock OSF https://osf.io/mdnwg; 2023.

\bibitem[{Rogachev and Sysoeva(2024)Rogachev, Anton and Sysoeva, Olga}]{42}
Rogachev A, Sysoeva O.
\newblock Neural tracking of natural speech in children in relation to their receptive speech abilities.
\newblock Cognitive Systems Research 2024;86:101236.

\bibitem[{Bianco et~al.(2024)Bianco, Roberta and Zuk, Nathaniel J and Bigand, F{'e}lix and Quarta, Eros and Grasso, Stefano and Arnese, Flavia and Ravignani, Andrea and Battaglia-Mayer, Alexandra and Novembre, Giacomo}]{43}
Bianco R, Zuk NJ, Bigand F, Quarta E, Grasso S, Arnese F, et~al.
\newblock Neural encoding of musical expectations in a non-human primate.
\newblock Current Biology 2024;34(2):444--450. e5.

\bibitem[{Mischler et~al.(2023)Mischler, Gavin and Raghavan, Vinay and Keshishian, Menoua and Mesgarani, Nima}]{44}
Mischler G, Raghavan V, Keshishian M, Mesgarani N.
\newblock naplib-python: Neural Acoustic Data Processing and Analysis Tools in Python.
\newblock arXiv preprint arXiv:230401799 2023;.

\bibitem[{Christian et~al.(2023)Christian, Brodbeck and Proloy, Das and Teon L, Brooks and Samir, Reddigari and J, Kulasignham.}]{45}
Christian B, Proloy D, Teon~L B, Samir R, J K, EelBrain-toolkit; 2023.

\bibitem[{Gramfort et~al.(2013)Gramfort, Alexandre and Luessi, Martin and Larson, Eric and Engemann, Denis and Strohmeier, Daniel and Brodbeck, Christian and Goj, Roman and Jas, Mainak and Brooks, Teon and Parkkonen, Lauri and H{\"a}m{\"a}l{\"a}inen, Matti}]{46}
Gramfort A, Luessi M, Larson E, Engemann D, Strohmeier D, Brodbeck C, et~al.
\newblock MEG and EEG data analysis with MNE-Python.
\newblock Frontiers in Neuroscience 2013;7.
\newblock \urlprefix\url{https://www.frontiersin.org/articles/10.3389/fnins.2013.00267}.

\bibitem[{Delorme and Makeig(2004)Delorme, A. and Makeig, S.}]{47}
Delorme A, Makeig S.
\newblock EEGLAB: an open source toolbox for analysis of single-trial EEG dynamics including independent component analysis.
\newblock J Neurosci Methods 2004;134(1):9--21.

\bibitem[{Alain(2023)Alain, de Cheveign{'e}.}]{48}
Alain dC, NoiseTools; 2023.
\newblock \urlprefix\url{http://audition.ens.fr/adc/NoiseTools/}.

\bibitem[{Bialas and Dou(2023)Bialas, Ole and Dou, Jin}]{49}
Bialas O, Dou J, mTRFpy; 2023.

\bibitem[{Ehinger and Dimigen(2019)Ehinger, Benedikt V and Dimigen, Olaf}]{51}
Ehinger BV, Dimigen O.
\newblock Unfold: an integrated toolbox for overlap correction, non-linear modeling, and regression-based EEG analysis.
\newblock PeerJ 2019;7:e7838.

\bibitem[{Di~Liberto et~al.(2023)Di Liberto, Giovanni M. and Crosse, Michael J. and Zuk, Nathaniel J. and Nidiffer, Aaron R. and Haro, Stephanie and Cantisani, Giorgia}]{52}
Di~Liberto GM, Crosse MJ, Zuk NJ, Nidiffer AR, Haro S, Cantisani G, CNSP resources; 2023.
\newblock \urlprefix\url{https://github.com/CNSP-Workshop/CNSP-resources Commit ID: 714e044934c94da1c0fc175513ca0952f22a9daa}.

\bibitem[{Di~Liberto et~al.(2019)Di Liberto, Giovanni M. and Wong, Daniel and Melnik, Gerda Ana and de Cheveigne, Alain}]{53}
Di~Liberto GM, Wong D, Melnik GA, de~Cheveigne A.
\newblock Low-frequency cortical responses to natural speech reflect probabilistic phonotactics.
\newblock NeuroImage 2019;196:237--247.

\bibitem[{Di~Liberto et~al.(2021)Di Liberto, Giovanni M. and Nie, Jingping and Yeaton, Jeremy and Khalighinejad, Bahar and Shamma, Shihab A. and Mesgarani, Nima}]{54}
Di~Liberto GM, Nie J, Yeaton J, Khalighinejad B, Shamma SA, Mesgarani N.
\newblock Neural representation of linguistic feature hierarchy reflects second-language proficiency.
\newblock NeuroImage 2021;227:117586--117586.

\bibitem[{Teoh et~al.(2019)Teoh, Emily S. and Cappelloni, Madeline S. and Lalor, Edmund C.}]{55}
Teoh ES, Cappelloni MS, Lalor EC.
\newblock Prosodic pitch processing is represented in delta-band EEG and is dissociable from the cortical tracking of other acoustic and phonetic features.
\newblock European Journal of Neuroscience 2019;50(11):3831--3842.
\newblock \urlprefix\url{https://doi.org/10.1111/ejn.14510}.

\bibitem[{Gillis et~al.(2023)Gillis, M. and Vanthornhout, J. and Francart, T.}]{56}
Gillis M, Vanthornhout J, Francart T.
\newblock Heard or Understood? Neural Tracking of Language Features in a Comprehensible Story, an Incomprehensible Story and a Word List.
\newblock eNeuro 2023;10(7).

\bibitem[{Verschueren et~al.(2022)Verschueren, Eline and Gillis, Marlies and Decruy, Lien and Vanthornhout, Jonas and Francart, Tom}]{57}
Verschueren E, Gillis M, Decruy L, Vanthornhout J, Francart T.
\newblock Speech Understanding Oppositely Affects Acoustic and Linguistic Neural Tracking in a Speech Rate Manipulation Paradigm.
\newblock The Journal of Neuroscience 2022;42(39):7442--7453.
\newblock \urlprefix\url{https://www.jneurosci.org/content/jneuro/42/39/7442.full.pdf}.

\bibitem[{Lesenfants et~al.(2019)Lesenfants, Damien and Vanthornhout, Jonas and Verschueren, Eline and Francart, Tom}]{58}
Lesenfants D, Vanthornhout J, Verschueren E, Francart T.
\newblock Data-driven spatial filtering for improved measurement of cortical tracking of multiple representations of speech.
\newblock Journal of Neural Engineering 2019;.

\bibitem[{Karunathilake et~al.(2023)Karunathilake, IM Dushyanthi and Dunlap, Jason L and Perera, Janani and Presacco, Alessandro and Decruy, Lien and Anderson, Samira and Kuchinsky, Stefanie E and Simon, Jonathan Z}]{59}
Karunathilake ID, Dunlap JL, Perera J, Presacco A, Decruy L, Anderson S, et~al.
\newblock Effects of aging on cortical representations of continuous speech.
\newblock Journal of neurophysiology 2023;129(6):1359--1377.

\bibitem[{de~Cheveign{'e} and Nelken(2019)de Cheveign{'e}, Alain and Nelken, Israel}]{60}
de~Cheveign{'e} A, Nelken I.
\newblock Filters: when, why, and how (not) to use them.
\newblock Neuron 2019;102(2):280--293.

\bibitem[{Marion et~al.(2021)Marion, Guilhem and Di Liberto, Giovanni M. and Shamma, Shihab A.}]{61}
Marion G, Di~Liberto GM, Shamma SA.
\newblock The Music of Silence. Part I: Responses to Musical Imagery Accurately Encode Melodic Expectations and Acoustics.
\newblock Journal of Neuroscience 2021;.

\bibitem[{Chalehchaleh et~al.(2024)Chalehchaleh, Amirhossein and Winchester, Martin Methven and Di Liberto, Giovanni}]{62}
Chalehchaleh A, Winchester MM, Di~Liberto G.
\newblock Robust assessment of the cortical encoding of word-level expectations using the temporal response function.
\newblock bioRxiv 2024;p. 2024.04. 03.587931.

\bibitem[{Wilkinson et~al.(2016)Wilkinson, Mark D and Dumontier, Michel and Aalbersberg, IJsbrand Jan and Appleton, Gabrielle and Axton, Myles and Baak, Arie and Blomberg, Niklas and Boiten, Jan-Willem and da Silva Santos, Luiz Bonino and Bourne, Philip E}]{63}
Wilkinson MD, Dumontier M, Aalbersberg IJ, Appleton G, Axton M, Baak A, et~al.
\newblock The FAIR Guiding Principles for scientific data management and stewardship.
\newblock Scientific data 2016;3(1):1--9.

\bibitem[{Dmochowski et~al.(2018)Dmochowski, Jacek P and Ki, Jason J and DeGuzman, Paul and Sajda, Paul and Parra, Lucas C}]{64}
Dmochowski JP, Ki JJ, DeGuzman P, Sajda P, Parra LC.
\newblock Extracting multidimensional stimulus-response correlations using hybrid encoding-decoding of neural activity.
\newblock NeuroImage 2018;180:134--146.

\bibitem[{Di~Liberto et~al.(2021)Di Liberto, Giovanni M and Barsotti, Michele and Vecchiato, Giovanni and Ambeck-Madsen, Jonas and Del Vecchio, Maria and Avanzini, Pietro and Ascari, Luca}]{65}
Di~Liberto GM, Barsotti M, Vecchiato G, Ambeck-Madsen J, Del~Vecchio M, Avanzini P, et~al.
\newblock Robust anticipation of continuous steering actions from electroencephalographic data during simulated driving.
\newblock Scientific reports 2021;11(1):23383.

\end{thebibliography}



\end{document}